\documentclass[sn-basic]{sn-jnl}

\jyear{2021}%

\theoremstyle{thmstyleone}%
\theoremstyle{thmstyletwo}%

\theoremstyle{thmstylethree}%
\newcommand{\cii}{[C{\footnotesize II}]}

\usepackage{tablefootnote}
\usepackage[symbol]{footmisc}
\usepackage{lscape}

\usepackage{comment}
\begin{document}

\title[Small Molecules, Big Impact]{\centering Small Molecules, Big Impact: \\A tale of hydrides past, present, and future}


\author*[1]{\fnm{Arshia M.} \sur{Jacob}}\email{ajacob51@jhu.edu}

\affil*[1]{\orgdiv{William H. Miller III Department of Physics \& Astronomy}, \orgname{Johns Hopkins University}, \orgaddress{\street{3400 N. Charles Street}, \city{Baltimore}, \postcode{21218}, \state{Maryland}, \country{USA}}}




\abstract{Formed at an early stage of gas-phase ion-molecule chemistry, hydrides -- molecules containing a heavy element covalently bonded to one or more hydrogen atoms -- play an important role in interstellar chemistry as they are the progenitors of larger and more complex species in the interstellar medium. In recent years, the careful analysis of the spectral signatures of hydrides have led to their use as tracers of different constituents, and phases of the interstellar medium and in particular the more diffuse environments. Diffuse clouds form an essential link in the stellar gas life-cycle as they connect both the late and early stages of stellar evolution. As a result, diffuse clouds are continuously replenished by material which makes them reservoirs for heavy elements and hence ideal laboratories for the study of astrochemistry. This review will journey through a renaissance of hydride observations detailing puzzling hydride discoveries and chemical mysteries with special focus carbon-bearing hydrides to demonstrate the big impact of these small molecules and ending with remarks on the future of their studies.}




\maketitle

\section{Introduction}\label{sec:Introduction}
\begin{figure}
    \centering
    \includegraphics[width=\textwidth]{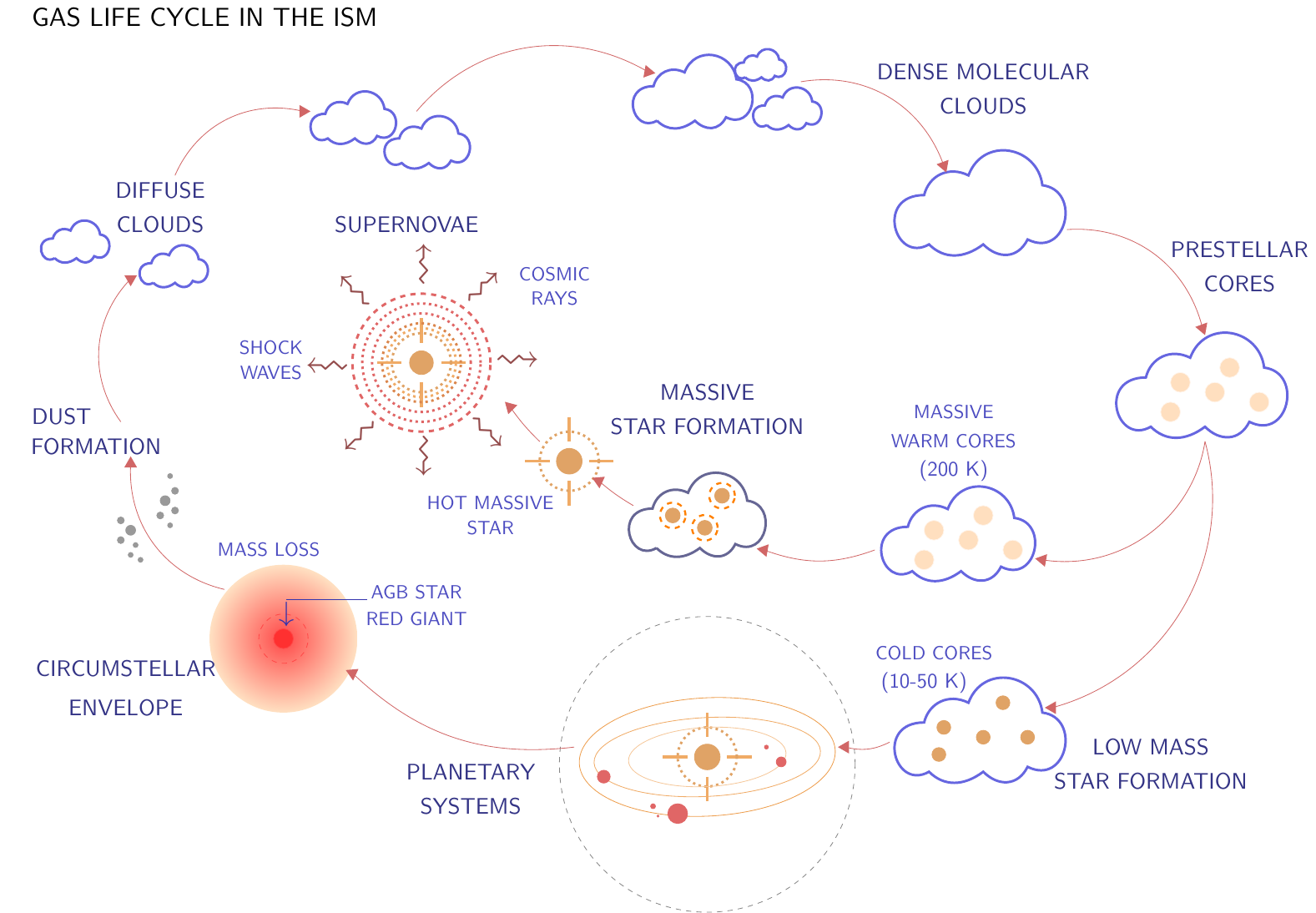}
    \caption{The life cycle of stars, gas and dust in the ISM. Adapted from an illustration by Steven Simpson \citep[][Sky \& Telescope Magazine]{Verschuur1992} with permission.}
    \label{fig:star-gas-life_cycle}
\end{figure}

Throughout much of human history, the space between stars collectively known as the interstellar medium (ISM) was thought to be empty. It was only toward the end of the 19$^{\rm th}$ century that scientists began to discover the existence of a gaseous ISM, with the first direct evidence provided by the discovery of spectral lines in the optical spectra of emission nebulae, such as the Orion Nebula by \citet{Huggins1864} and \citet{Huggins1889}. We now know that the ISM is highly complex and inhomogeneous, consisting of different phases composed of gas and dust in the form of dense clouds (the birthplace of stars) or more diffuse gas permeated by energetic particles, exposed to intense radiation fields and violent shocks. Broadly these different phases are classified based on variations in the ionisation state of hydrogen (existing in either its neutral atomic, ionized or molecular forms),  and the physical conditions prevalent in the region such as its gas temperature and density \citep[see for example,][]{Snow2006, Draine2011}. The earliest of many models characterising the multi-phase ISM was put forth by \citet{Field1969}, who postulated the existence of a two phase ISM consisting of a cold neutral medium (CNM) and a warm neutral medium (WNM), respectively, by assuming that the atomic gas in the ISM is in thermal equilibrium. An additional phase was invoked by \citet{McKee1977}, known as the hot ionised medium (HIM) constituting ‘coronal’ gas with temperatures in excess of 10$^6$~K. This phase is primarily formed from ionised bubbles and traced by X-ray emission. The multi-phase schematic of the ISM was later extended to include the warm ionised medium (WIM), to account for a more widespread component comprising of material that is predominantly ionised by radiation from massive stars. While the warm and hot ionised phases fill a significant volume fraction they only represent a small fraction of the total gas mass, the bulk of which is contained in the cooler and denser \textit{neutral} phases of the ISM. Therefore, the neutral ISM represents an enthralling laboratory that facilitates the study of a wide range of astrophysical phenomena, as it forms a reservoir of material from which stars form, linking the physics between stellar and galactic scales \citep[see][for a comprehensive review]{Saintonge2022}, as illustrated in Fig.~\ref{fig:star-gas-life_cycle}. 

Being composed of only trace abundances of elements heavier than helium, at first glance, the ISM presents a hostile environment for the formation, survival and growth of molecules and molecular complexity. However, a plethora of molecules ranging from simple diatomic species to fullerenes containing up to 60 C-atoms \citep{Sellgren2010} have been identified in a variety of astronomical objects ranging from comets and exoplanet atmospheres to interstellar and circumstellar environments and external galaxies. This has resulted in the successful identification of a total of 260 molecules over the past 80~years and as of this review \citep[see][for a census]{McGuire2022}. These exciting discoveries have been made possible in large thanks to developments in receiver technology particularly in the field of radio astronomy and advancements in laboratory spectroscopy. A multidisciplinary science, the field of astrochemistry has evolved to encompass topics of interest to astronomers, physicists, and chemists, alike.
\begin{center}
    \textit{But what do we learn from these molecules?}\\
\end{center} 

Owing to the complexity of their energy level structures and as a result their chemical fingerprints or spectra-- observed in either emission or absorption, molecules are unique diagnostic probes of the physical conditions in which they are found. In addition to forming excellent probes of astronomical conditions, molecules play a central role in influencing the thermal and ionisation structure of gas, initiating condensations and instabilities responsible for the formation of stars and as a result regulate the evolution of the environments in which they are formed. While astronomical spectra reveal a wealth of knowledge, their interpretation is complicated by the need for a detailed understanding of their chemistry aided by both theoretical calculations and laboratory measurements which are often limited to the study of simpler systems. To that end, small molecules whose chemistry is well understood and which form the fundamental building blocks from which increasingly more complex species are formed, show the greatest promise for extending our understanding of the chemical Universe. This review focuses on such a bottom-up approach, with particular emphasis on the use of hydrides (molecules or molecular ions of the form $X$H$_{\rm n}$ and $X$H$^{+}_{\rm n}$) in investigating the properties of the diffuse and translucent gas phases of the neutral ISM. \\

\subsection{A brief history of hydride observations}\label{subsec:hydride_history}
From a historical perspective, hydrides were the first gas-phase molecules to be detected in the ISM. \citet{Swings1937} identified the first interstellar molecule -- the methylidyne radical, CH, via one of the transitions between its A$^{2}\Delta$-X$^2\Pi$ electronic sates near 4300.2~\AA\, based on optical absorption spectra observed toward several early type stars by \citet{Dunham1937} \citep[and later][]{Adams1941}. This discovery was later confirmed by the subsequent identification of CH lines near 3880~\AA\ corresponding to its B$^{2}\Sigma$-X$^2\Pi$ electronic transitions by \citet{McKellar1940}. Motivated by this discovery, careful laboratory investigations of the other unidentified absorption line bands lying close to 3874.5~\AA\ and 4232.6~\AA\ by \citet{McKellar1940} and \citet{Douglas1941, Douglas1942}, led to the subsequent identification of CN and CH$^+$, respectively, in interstellar space, marking the birth of molecular astrophysics. 

Akin to the optical regime, hydride signatures became the first molecular lines to be detected at radio wavelengths through the ground-state 18~cm transitions of OH by \citet{Weinreb1963}. Split into four hyperfine structure (HFS) lines between 1665~MHz and 1620~MHz, early observations of the OH ground-state transitions revealed line intensities that differed from that predicted theoretically. These observed enhancements were interpreted as being caused by the amplification of radiation by stimulated emission and led to the detection of the first interstellar maser \citep{Weaver1965Natur}. The detection of OH motivated the search for polyatomic species at radio wavelengths which resulted in the detection of two more hydrides, NH$_3$ \citep{Cheung1968} and H$_2$O \citep{Cheung1969Natur} and soon even larger species like H$_2$CO \citep{Snyder1969}. 

\begin{figure}
\centering 
    \includegraphics[width=1\textwidth]{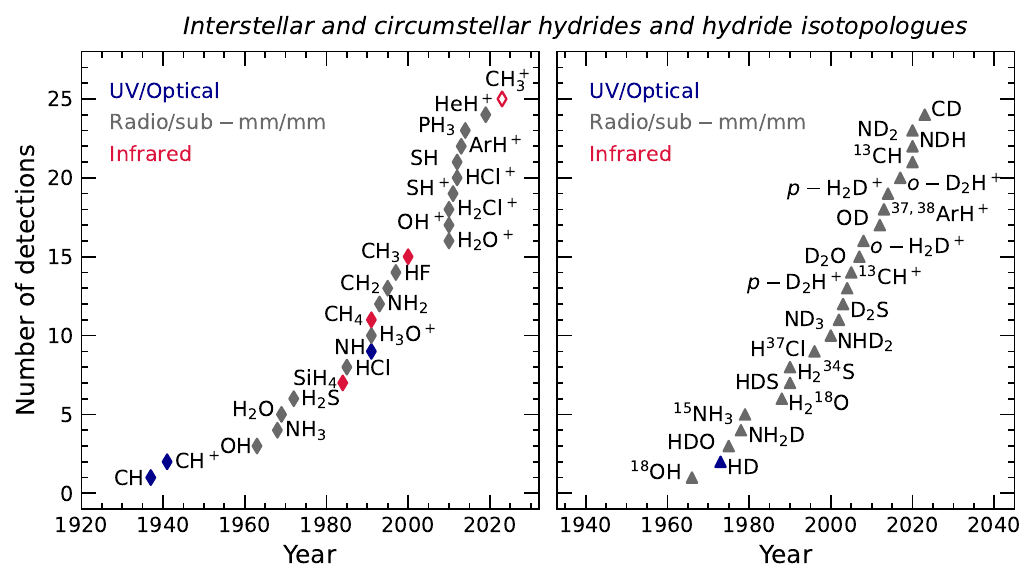}
    \caption{Census of all astrophysical detections of hydrides to date (filled diamonds, left-hand panel) and hydride isotopologues (filled triangles, right-hand panel). Hydrides detected for the first time via their electronic transitions at UV/optical wavelengths are marked in blue, those observed through their rotational transitions at radio/sub-mm/mm wavelengths are displayed in grey while those observed through transitions between vibrational states at infrared wavelengths are displayed in red. Unfilled markers represent tentative detections.}
    \label{fig:hydride_history}
\end{figure}

A few years later, the HFS lines between the $\Lambda$-doublet levels of the rotational ground state of CH lying at radio wavelengths (9~cm or 3.3~GHz) was detected by \citet{Rydbeck1973}. However, similar to OH, the intensities of the three ground state HFS lines of CH were found to be inconsistent with predictions made under conditions of local thermodynamic equilibrium (LTE). Although much weaker than the observed OH maser emission, the anomalous excitation of CH was qualitatively understood by assuming a pumping cycle involving collisional excitation processes similar to that of OH \citep{Bertojo1976, Elitzur1977}. However, the relative intensities of the CH HFS lines and in particular the dominance of the lowest frequency line was not well understood and this greatly limited its use as a tracer of molecular gas in the ISM. While a number of other light hydrides were detected in the early days of radio astronomy (see Fig.~\ref{fig:hydride_history}), developments in this field garnered interests in observing increasingly more complex molecules and exploring the chemical richness of interstellar environments. Furthermore, due to their small size, hydrides are characterised by widely spaced energy levels especially between their energetically lowest rotational states. Therefore, the fundamental rotational transitions of most light hydrides lie at higher frequencies. Hence, all subsequent detections of hydrides have come with advancements in heterodyne receiver technology and the new access it provided to the millimetre (mm)/sub-millimetre (sub-mm) and far-infrared (FIR) wavelength regimes. Largely carried out using the Herschel Space Observatory \citep[HSO;][]{Pilbratt2010}, the Atacama Pathfinder EXperiment \citep[APEX;][]{Gusten2006} 12~m telescope, and the Stratospheric Observatory for Infrared Astronomy \citep[SOFIA;][]{Young2012}, these observations have roughly doubled the number of hydrides (and hydride isotopologues) detected in space (see Fig~\ref{fig:hydride_history}). The rotational transitions of hydrides are generally observed in widespread absorption along sight lines toward bright background continuum sources yielding robust measurements of column densities across Galactic scales unlike absorption spectroscopy at optical wavelengths which is limited to nearby (a few kpc) and bright sources. This has resulted in renewed interests in the study of interstellar hydrides over Galactic and extragalactic scales. Studies, which have subsequently established the use of specific hydrides as sensitive tracers of the physics and chemistry at play in different phases of the ISM \citep[refer][for a detailed review]{Gerin2016}.

\subsection{Scope and structure of this review}
The main objective of this review is to emphasise the diagnostic properties of light carbon-bearing hydrides, CH and CH$_2$. This will be achieved by addressing the following problems: 
\begin{itemize}
    \item The column densities of molecular hydrogen have historically been estimated using that of CO as proxy, as H$_2$ is practically invisible under the quiescent conditions of the ISM due to its weak dipole moment. However, mounting evidence indicates the presence of a molecular gas component in the absence of CO. Section~\ref{subsec:CH_CO-darkgas} details how reliably hydrides and in particular CH is in tracing this missing gas component across Galactic scales.

    \item As discussed in Sect.~\ref{subsec:hydride_history}, despite the relative ease with which it is observed, the diagnostic properties of the ground state radio lines of CH at 3.3~GHz have not been fully exploited because the excitation of these lines are not well understood. Section~\ref{subsec:CH_maser} presents a solution to this longstanding mystery by combing recently computed collisional rate coefficients with new constraints provided by multi-wavelength observations of CH.  
    \item Despite showing substantial deviations from elemental abundances caused by isotope selective effects, the isotopic abundances estimated using interstellar molecules measured over large scales form essential diagnostics of the history of Galactic nucleosynthesis and chemical enrichment. 
    As an important constituent of gas-phase interstellar chemistry, CH and its isotopologues will not only provide potentially unbiased estimates of the isotopic abundances but will also provide a great deal of information about critical isotope exchange reactions. Section~\ref{subsec:CH_isotopes} describes the search for, and utility of the CH isotopologues -- $^{13}$CH and CD. 
    \item Theoretically, CH$_2$ was predicted to arise in diffuse clouds spatially coincident with CH \citep{prasad1980model, van1986comprehensive}. This is not surprising considering the chemical association between these two species. Contrary to these predictions, the initial detection of CH$_2$ emission in the ISM was made towards dense, hot cores such as Orion~KL. However, subsequent observations toward the same regions, conducted using larger telescopes, have failed to detect CH$_2$, raising questions about the source and nature of its emission. With the aid of new observations, the elusive nature of this species is addressed in Section.~\ref{sec:elusive_ch2}. 
    
\end{itemize}
Section~\ref{sec:the_present} extends the discussion to current and on-going hydride research and finally Sect.~\ref{sec:summary_outlook} ends with an outlook on the exciting future of hydride studies in the coming years. The relevant spectroscopic parameters and frequencies of the transitions discussed in this review are tabulated in Appendix~\ref{appendix:frequencies}.

\section{Probing the neutral ISM}\label{sec:neutral_ISM}

Considered neutral with respect to the ionisation state of hydrogen, the neutral phases of the ISM are primarily composed of atomic (H{\small I}) and molecular (H$_2$) hydrogen, as most of its volume lacks Lyman continuum photons (with energies $>$13.6~eV) capable of ionising it. However, this phase is not strictly neutral because species with ionisation potentials lower than that of hydrogen can still be ionised, for example carbon with a first ionisation potential of 11.26~eV. In addition, cosmic rays can freely penetrate and (partially) ionise gas in dense clouds that are well shielded from other sources of ionisation like UV and X-ray photons, enabling exothermic ion-neutral reactions. Furthermore, as the major component of the neutral ISM, cloud formation must involve a phase transition from gas that is dominated by H{\small I} gas to that which is dominated by H$_2$, which then fuels the formation of stars. Therefore investigating the neutral ISM is critical for our understanding of molecular cloud formation -- important stellar nurseries and chemical laboratories, responsible for initiating the growth of chemical complexity in the ISM both locally and across Galactic scales. 

The serendipitous discovery of radio waves by Jansky and pioneering efforts by \citet{Reber1944} and others, along with advancements in technology led to the birth of radio astronomy in the early 1940s, and with it the detection of H{\small I} emission at 21~cm ($\nu = 1.420405~$GHz) by \citet{Ewen1951}. Providing new access to the radio skies and unaffected by dust, the H{\small I} 21~cm line has, since its discovery been widely used as a powerful tool to investigate the atomic gas reservoir of galaxies, their kinematics, and large scale structure. Astronomical signatures of H$_2$ were first detected in the Galactic ISM via Lyman resonance absorption bands between 1000 ~\AA\ and 1100~\AA\ by \citet{Carruthers1970} using the Aerobee-150 rocket toward luminous stars $\epsilon$ Persei and $\xi$ Persei. While searches for cosmic H$_2$ at different wavelengths followed, aided by advancements in detector technology, theory and laboratory experiments \citep[see review by][]{Williams1999}, observations of H$_2$-- the most abundant molecular species in the Universe, has traditionally been difficult.

\subsection{Challenges in observing molecular hydrogen}\label{ref:observing_H2}

Spectroscopic studies of the electronic bands, pure and highly excited quasibound rotational transitions as well as forbidden vibrational transitions of molecular hydrogen, have been carried out in the laboratory using a variety of methods over the years. The techniques employed range from Raman spectroscopy to advanced high-precision measurements involving narrowband vacuum ultraviolet lasers (VUV), Fourier-transform spectrometers, and synchrotron radiation
sources \citep[for example,][]{Stoicheff1957, Herzberg1959, Cooper1970, Herzberg1972, Bragg1982, Jennings1982, Jennings1983, Jennings1984, Chupka1987, Ubachs2004, Reinhold2006}. However, because the physical conditions associated with the astrophysical environments in which H$_2$ lines emit, may not be reproducible in the laboratory, measurements of the H$_2$ spectra are (mostly) limited to lower rotational levels ($J<29$). Therefore, laboratory efforts were often both accompanied by, and stimulated theoretical predictions of the H$_2$ spectrum which entailed detailed fundamental ab initio calculations, with early contributions from \citet{Kolos1964}, \cite{Wolniewicz1966}, \citet{Kolos1968}, and \citet{Abgrall1993}, which were then refined by the inclusion of non-adiabatic, relativistic, and quantum-electrodynamical (QED) effects by \citet{Pachucki2009} and \citet{Komasa2011}. Most notably, by taking advantage of the recent theoretical ab initio calculations, \citet{Roueff2019} computed the full infrared spectrum and the emission probabilities of the electric quadrupole and magnetic dipole transitions present within the ground electronic state of molecular hydrogen with an unprecedented accuracy. 

Observationally, because H$_2$ is a homonuclear and symmetric molecule, it does not possess a permanent dipole moment and its excited ro-vibrational states only radiate through a weak quadrupole transition. In addition, the H$_2$ molecule has a large rotational constant owing to its low moment of inertia, which results in widely spaced energy levels with its first rotational transition lying at 28~$\mu$m with an upper level energy, $E_{\rm u}/k_{\rm B}$, of $\sim\!500~$K\footnote{Now accessible with the Mid-Infrared Instrument on the James Webb Space Telescope (JWST) \citep{Rieke2015} after the end of the Spitzer mission.}. 
 The ro-vibrational bands of H$_2$ at mid-, and near-infrared wavelengths, whilst widely observed are only excited under special conditions, either by collisions in interstellar shock waves or by UV resonance \citep{Hartigan1989, Garden1990, Burton1992, Neufeld2006, Habart2011, Yuan2011, Santangelo2014, Neufeld2019}. Therefore, without any transitions at radio wavelengths, observationally we lack a direct probe of the coldest and densest parts of the ISM as H$_2$ cannot be excited in the cooler ISM gas phases associated with star formation. 
 
In absorption, the electronic transitions of H$_2$ observed at far-UV wavelengths are restricted to nearby bright hot stars and toward bright quasars \citep{Savage1977, Rachford2002, Rachford2009, Pan2005, Burgh2007, Noterdaeme2010, Balashev2014, Ubachs2016, Shull2021}. While the strongest electric dipole-allowed transitions for H$_2$ lie at rest frame wavelengths of 912--1155~\AA\ in the Lyman and Werner bands, for high redshift systems at $z>2$ the absorption lines shift such that their transmission approaches visible wavelengths ($\lambda>3050$~\AA) \citep{Ledoux2003, Noterdaeme2008, Noterdaeme2010}. The energy spectra of these bands were first measured by \citet{Dabrowski1984} using experimental setups involving VUV absorption and emission flash discharge spectra. \\

The challenges involved in observing H$_2$ in cooler environments of the ISM are overcome by using other associated (or co-spatial) chemical species including dust, as surrogates. This is feasible because of the specific physical and chemical conditions pertinent to the formation of interstellar H$_2$ \citep[see][for a detailed overview]{Vidali2013, Wakelam2017}. Most of the H$_2$ in the ISM (with the exception of that formed in the early Universe), has long been thought to be formed via catalytic reactions on the surfaces of interstellar dust grains since gas-phase reactions are inefficient in predicting its abundances \citep{Gould1963, Hollenbach1971}. Similarly, the formation of other interstellar molecules have also been shown to result from chemical-exchange reactions between atoms that are chemically bound to interstellar grains and those in the gas-phase. As a result, the most commonly used proxy for measuring the total molecular reservoir in both local \citep{Leroy2011, Saintonge2017} and high redshift galaxies \citep{Pavesi2018, Tacconi2018} is the $J=1-0$ transition of CO -- the second most abundant molecule in the ISM -- at 2.6~mm (115.27~GHz). The integrated intensity of CO is scaled to that of H$_2$ by an empirical CO-to-H$_2$ conversion factor, $X$(CO). The $X$(CO) factor has a commonly adopted value of 2$\times10^{20}~$cm$^{-2}$(K~km~s$^{-1}$)$^{-1}$ in studies of the Milky Way, with an uncertainty of 30\% \footnote{The uncertainties in the computed values of the $X$(CO) factor in both the Milky Way and external galaxies \citep{Sandstrom2013}, and the validity and assumptions involved in prescribing to a single value are discussed thoroughly in the review by \cite{Bolatto2013}.} \citep{Bolatto2013}. 
Broadly, the combination of neutral hydrogen traced by the H{\small I} 21~cm line and molecular gas traced using the $J=1-0$ transition of CO can be used to account for the total neutral hydrogen column density. \textit{But a key question that arises is, how reliable this combination is, in reproducing the total neutral gas reservoirs present in different galaxies?}

\subsection{Emergence of a new phase of the ISM-- CO-dark molecular gas}\label{subsec:CO-dark_gas}
The ability of the column densities determined from the H{\small I} 21~cm line and H$_2$ scaled using the $X$(CO) factor to trace the total neutral gas column density is assessed by comparing these values with large scale, spectrally resolved surveys made using other tracers of the neutral gas column such as the diffuse gamma-ray flux \citep[EGRET;][]{Grenier2005} and the sub-mm dust optical depth \cite[Planck;][]{Planck2011}. An excess of gas invisible in emission from H{\small I} and CO was inferred from these comparisons \citep[see for example Fig.~1 of][]{Grenier2005}. This excess gas can be attributed to either the saturation of the H{\small I} emission line profile or the inability of the standard CO proxy to trace H$_2$, in regions of low dust column densities (or visual extinction, $A_{\upsilon} = 0.05$--2~mag). This is particularly the case for low-metallicity galaxies \citep{Schruba2012}, where the lower dust abundances permit the penetration of far-UV photons deeper into the molecular cloud in comparison to more metal rich environments, such that $\sim\!70\,$\% of the molecular gas mass is not traced by CO. The presence of a similar excess gas component was previously revealed by studies comparing 100~$\mu$m dust emission from the Infrared Astronomical Satellite (IRAS) with observations of H{\small I} and CO toward high latitude cirrus clouds \citep{Reach1994, Meyerdierks1996, Boulanger1998A}. While the main constituent of this gas is sightline dependent, studies have shown that the excess dark gas is primarily molecular \citep{Liszt2018, Liszt2023}. The atomic gas content that is unaccounted for, can be understood using H{\small I} absorption measurements \citep{Li2003}, while the molecular gas content missed by CO requires the use of additional molecular gas tracers. \\

In spite of probing the bulk of the molecular material in the ISM, being less efficiently self-shielded than H$_2$ in warmer and more diffuse regions, CO is readily photodissociated by the surrounding interstellar radiation field. Therefore, in regions exposed to UV irradiation where carbon primarily resides in either its atomic or ionised forms, a substantial amount of molecular hydrogen (often containing a mass equivalent to that of the atomic gas component) is left undetected and dubbed as `CO-dark' molecular gas. Since knowledge of the physical and excitation properties of dark molecular gas is vital for our understanding of the chemical evolution and diversity of the ISM, it is of great importance to investigate the nature and composition of this new gas phase. Moreover, the presence of CO-dark molecular gas leads to uncertainties in the total molecular gas mass which in turn limits our understanding of the efficiency with which gas is converted into stars.

A chemical precursor to CO and widely observed within the diffuse and translucent regions of the ISM, the 158~$\mu$m cooling line of C$^+$ has been extensively used to study the properties of CO-dark gas in both the Milky Way and nearby galaxies 
\citep{Pineda2013, Langer2014, Bigiel2020, Chevance2020, Madden2020}. In recent times, the capabilities of the C$^+$ line to trace both normal, and CO-dark molecular gas has been extended to the high-redshift universe. This expansion has been made possible thanks to the accessibility of the the Atacama Large Millimeter Array (ALMA), to the sub-mm window \citep{Zanella2018, Bethermin2020, Schaerer2020, Tacconi2020}. The 158~$\mu$m C$^+$ line is also one of the main targets of the Galactic/Extragalactic Spectroscopic Terahertz \citep[GUSTO;][]{Goldsmith2022} balloon mission which will map large parts of the Galactic Centre and Large Magellanic cloud. 

While [C\,{\small II}] emission forms a convenient tool for quantifying the amount of CO-dark gas present in a galaxy, there are also limitations to its use \citep[see discussion in][]{Madden2020}. One important caveat is that the [C\,{\small II}] emission traces multiple phases of the ISM (both neutral and ionised) which deems it necessary to account for contributions from these varied phases before determining the total molecular gas reservoir. Galactic disk simulations of Milky Way-type galaxies, investigating the properties of CO-dark gas by \citet{Glover2016}, also question the utility of [C\,{\small II}] as a tracer of CO-dark H$_2$ gas given that the brightness of the [C\,{\small II}] emission is only weakly correlated with the H$_2$ surface density. Therefore, the amount of CO-dark H$_2$ gas being probed by [C\,{\small II}] emission must be compared with that traced by other species to better constrain the physical properties traced specifically by the CO-dark gas component. This can be achieved by either using observations of other ionised gas tracers like [N{\small II}] to calibrate contributions from the ionised phase (for example in surveys using the GUSTO mission) or by observing other molecular gas tracers including hydrides that are also tracers of a gas regime undetected by CO in emission.

\section{Hydrides as probes of diffuse and translucent molecular gas}
\begin{figure}
     \includegraphics[width=0.48\textwidth]{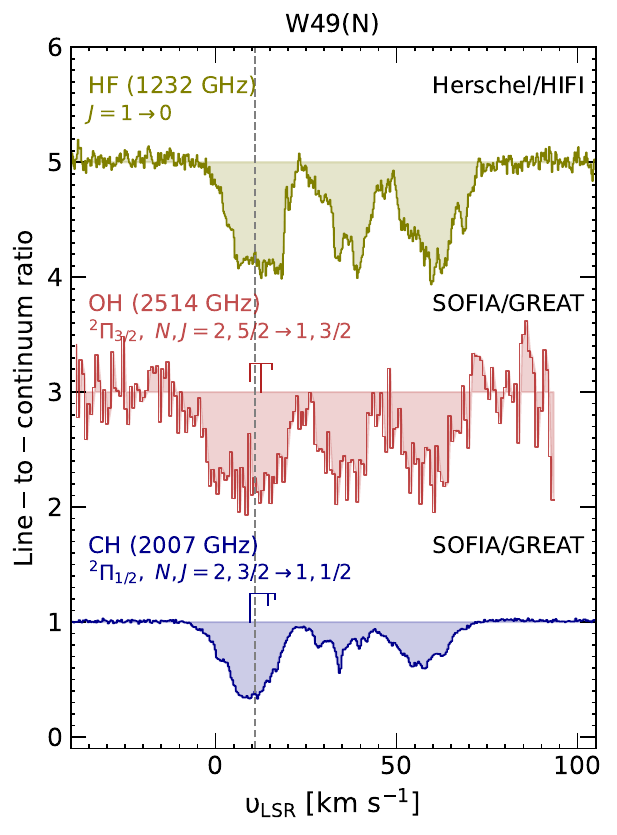}\quad
     \includegraphics[width=0.48\textwidth]{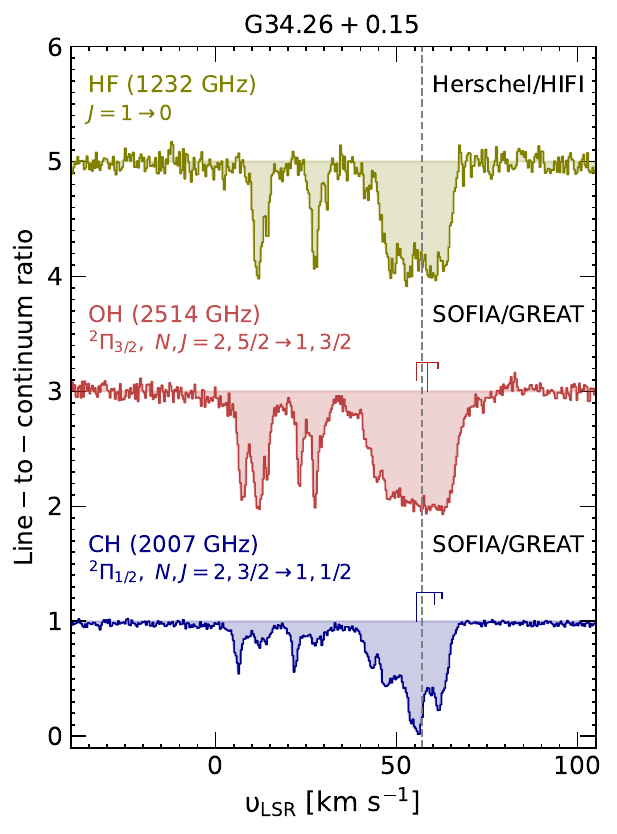}
     \caption{From top to bottom: Absorption spectra of HF $J=1\rightarrow 0$ at 1232~GHz, OH $^{2}\Pi_{3/2} N, J=2,5/2\rightarrow 1,3/2$ at 2514~GHz and CH $^{2}\Pi_{1/2} N, J=2,3/2\rightarrow 1,1/2$ at 2007~GHz in main-beam temperature scales normalised with respect to the continuum temperature, towards W49~(N) (left) and G34.26+0.15 (right). The HF and OH spectra are shifted along the y-axis for clarity and the relative intensities of the HFS-splitting components of OH and CH are marked above the spectra. The dashed vertical line indicates the systemic velocity of the background source. This figure is created based on data presented in \citet{Sonnentrucker2015}, \citet{Wiesemeyer2018} and \citet{Jacob2019}. }
     \label{fig:hydride_comparison}
 \end{figure}

In recent years, several hydrides such as CH \citep{Gerin2010, Wiesemeyer2018, Jacob2019}, OH \citep{Allen2015, Li2018, Rugel2018, Busch2019, Busch2021} and HF \citep{Neufeld2010, Sonnentrucker2015}, observed via their radio and FIR rotational transitions have been established as important tracers of CO-dark H$_2$ gas. While the radio ground-state transitions of OH and CH have been used as CO-dark gas proxies,  complexities in their excitation mechanisms make their utility challenging.
In contrast, the FIR rotational transitions of these hydride molecules possess high critical densities or high rates of spontaneous radiative decay, because of which they typically occupy their respective ground rotational states under diffuse cloud conditions. Furthermore, when observed toward bright background continuum sources their spectral line profiles are seen in pure absorption. Figure~\ref{fig:hydride_comparison} displays the observed absorption spectra of HF, OH and CH toward two luminous star-forming regions in the Milky Way, W49~(N) and G34.26+0.15. While the spectra of all three species are seen in widespread unsaturated absorption along the line-of-sight toward W49~(N) and G34.26+0.15, both HF and OH display saturated absorption line profiles, particularly at the systemic velocity of the background source and even in components along the line-of-sight toward G34.16+0.15. The optical depths\footnote{The optical depth, $\tau$, can be estimated from the ratio between the observed line brightness temperature ($T_{\rm l}$) and the background continuum temperature ($T_{\rm c}$) following the radiative transfer (RT) equation. In the case of absorption line spectroscopy, the RT equation can be approximated to $T_{\rm l} = T_{c\rm }{\rm e}^{-\tau}$.}, $\tau$, corresponding to these components approaches infinity as the line-to-continuum ratio tends toward a value of zero. In comparison, the absorption line profiles of CH are only marginally optically thick with $\tau_{\rm CH}<< 1$ in foreground absorbing material and up to a few (${\lesssim 3}$) for velocity components corresponding to the background source. 
The molecular column density, $N$, for a given species and transition is directly proportional to the optical depth \citep[see Eqn.~32 of][]{Mangum2015}, therefore, free from saturation and optical depth effects the determination of column densities from the CH radical appears to be much more straightforward. In addition, CH stands out from its hydride counterparts because it has a tight correlation with H$_2$ as established by \citet{Federman1982}, \citet{Mattila1986}, \citet{Sheffer2008} and \citet{Weselak2019}. For these reasons, in the following sections we detail the use of CH as a tracer for CO-dark molecular gas. 

\subsection{CH as a tracer of CO-dark \texorpdfstring{H$_2$}{H2} gas}\label{subsec:CH_CO-darkgas}
As discussed above, one of the primary reasons why CH has been used as a diffuse molecular gas tracer is based on its observed correlation with H$_2$, in gas with H$_2$ column densities between $10^{19}$~cm$^{-2}$ and $10^{21}$~cm$^{-2}$. This correlation, with a CH-H$_2$ relation of $3.5^{+2.1}_{-1.4}\times10^{-8}$, was first systematically established by \citet{Sheffer2008} using direct optical observations of both the A$^2\Delta$-X$^{2}\Pi$ systems of CH at 4300.2~\AA\ and bands of the Lyman B-X transitions of H$_2$ toward nearby stars. Akin to the optical observations of CH presented by these authors, the fundamental rotational transitions of CH connecting the $N,J = 1,3/2\rightarrow 1, 1/2$ \citep{Gerin2010, Qin2010, Bottinelli2014} and $N,J = 2,3/2\rightarrow 1, 1/2$ \citep{Wiesemeyer2018, Jacob2019} levels observed using Hereschel/HIFI and SOFIA/GREAT, respectively, are also seen in absorption but toward sightlines that probe molecular clouds across the Galaxy. The observed spectra of both sets of transitions show unsaturated absorption line profiles where that of the former (being easier to excite) displays a complex profile with a combination of emission from the background molecular cloud and absorption from both the surrounding envelope and foreground material while the spectra of the latter is seen in pure absorption. Nonetheless, the surrogate properties of CH have been used to both determine the canonical abundances of related species with respect to the total molecular gas, as well as evaluate their use as complementary tracers of diffuse molecular gas, like C$_2$H and HCO$^+$ \citep{Gerin2010, Liszt2023a}. \textit{However, is the correlation between CH and H$_2$, established by optical observations of the local diffuse ISM, valid over Galactic scales?}\\

\begin{figure}
    \centering
    \includegraphics[width=0.8\textwidth]{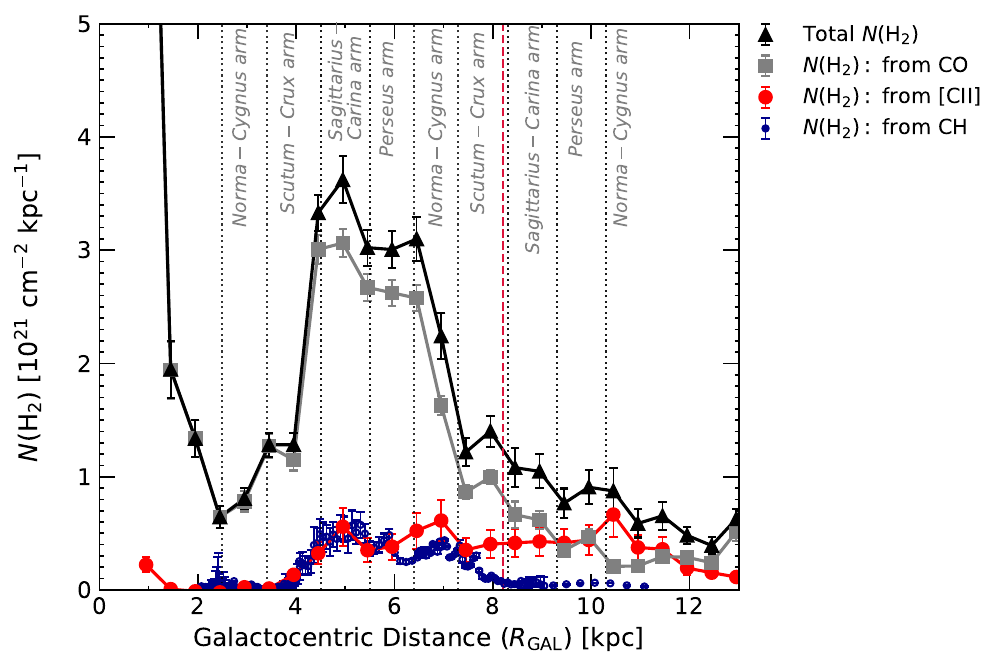}
    \caption{Radial distribution of the total H$_2$ column density in the plane of the Milky Way (black triangles) alongside the relative contributions of the H$_2$ column density traced by CO (grey squares) and CO-dark gas tracers like \cii\ (red circles) and CH (blue dots). The positions of the spiral arm crossings typically probed along the line-of-sight of a background continuum source located in the fourth quadrant are marked assuming spiral-arm widths of 0.85~kpc. The vertical dashed red line indicates the galactocentric distance to the Sun. Adapted with permission based on results presented in \citet{Pineda2013} and \citet{Jacob2019}.}
    \label{fig:CO-dark_gas_distribution}
\end{figure}

This can be assessed by benchmarking the use of CH with other widely used tracers of diffuse molecular and CO-dark gas. Although, it is a proficient tracer of different phases of the ISM, the 158~$\mu$m line emission of C$^+$ having been observed over 500 Galactic sightlines under the framework of the Galactic Observations of Terahertz C$^+$ \citep[GOT C+;][]{Pineda2013} survey forms an excellent test bed. \citet{Pineda2013} and \citet{Langer2014} decomposed contributions from the cold neutral component of the observed C$^+$ emissivity based on differences in gas densities traced by the different ISM phases. Figure~\ref{fig:CO-dark_gas_distribution} displays the azimuthally averaged radial distribution of H$_2$ column densities derived from the $N,J = 2,3/2\rightarrow 1, 1/2$ transitions of CH at 2~THz observed using SOFIA/upGREAT\footnote{SOFIA Science Mission Operations is jointly operated by the Universities Space Research Association, Inc., under NASA contract NAS2-97001, and the Deutsches SOFIA Institut under DLR contract 50 OK 0901 and 50 OK 1301 to the University of Stuttgart. upGREAT is financed by resources from the participating institutes, and by the Deutsche Forschungsgemeinschaft (DFG) within the grant for the Collaborative Research Centre 956, as well as by the Federal Ministry of Economics and Energy (BMWI) via the German Space Agency (DLR) under Grants 50 OK 1102, 50 OK 1103 and 50 OK 1104.} with that derived from \cii. Unsurprisingly (in spite of the smaller sample size of the CH observations) both distributions display a dual peak profile, peaking between 4 and 7.5~kpc (see Fig.~\ref{fig:CO-dark_gas_distribution}) or distances corresponding to the major spiral-arm intersections that contain the bulk of the Galactic material by mass. Moreover within the inner Galaxy at galactocentric distances, $R_{\rm GAL} < 8~$kpc, the H$_2$ column densities derived from both species, CH and C$^+$ resemble one another, lending credence to the use of CH as a tracer for H$_2$. At larger galactocentric distances ($>8$~kpc) however, the column densities of H$_2$ derived from CH are lower than those derived from \cii. This may in part be due to an observational bias as this analysis does not probe sufficient sightlines in the outer Galaxy or it may be due to the overall lower abundances of CH in the outer Galaxy as a result of reduced rates of star formation and lower elemental abundances present at these distances \citep{Gerin2015}. Therefore, an accurate interpretation at larger distances requires a larger sample of sight lines covering a range of Galactocentric distances. 

A noteworthy observation was made by Johansson (1979), who surveyed CH traced via the main line in its ground state $\Lambda$-doublet at 3.335~GHz across the Galaxy. Investigating trends in the radial distribution of CH number densities, \citet{Johansson1979} reported in their Fig.~11 a CH distribution that peaks between 5~kpc and 7~kpc. By further comparing the CH profiles with that of CO and H\,{\small I}, this author found that the number density of CH did not fall-off as rapidly, as in the case of CO at larger Galacto-centric radii ($>8~$kpc). Therefore, populating Fig.~\ref{fig:CO-dark_gas_distribution} with CH observations made toward larger distances will further our understanding of the CO-dark gas phase in the outer Galaxy. Recently, another CO-dark gas tracer and hydride, OH, has been detected across a thick disk of diffuse gas in the outer Galaxy ($R_{\rm GC}\sim 11$--12~kpc) by \citet{Busch2021}, which was previously undetected in CO emission.

Unfortunately, with the conclusion of the SOFIA mission, the opportunity to access the high-energy rotational lines of CH at terahertz wavelengths for extending such an analysis is no longer available (as of the time of writing this review). This redirects our attention towards other wavelengths to pursue similar investigations. Following \citet{Johansson1979}, one can carry out systematic observations of CH at radio wavelengths to calibrate the $X_{\rm CO}$ factor and study its use as a CO-dark gas tracer \citep{Magnani2005, Xu2016, Dailey2020}. However, large scale maps of the extended ISM have rarely been done using CH. This is attributed to the weaker spectral signal of CH when compared to more abundant species like CO, as well as due to the challenges involved in interpreting the ground state radio lines of CH, owing to their anomalous excitation and masing (see discussion in Sect.~\ref{subsec:hydride_history}). In the following section we delve into the complexities involved in the excitation of these lines and explore modelling techniques used to establish the utility of this line.

\subsection{Revisiting the anomalous excitation in the CH ground state}\label{subsec:CH_maser}
\begin{figure}
    \includegraphics[width=1\textwidth]{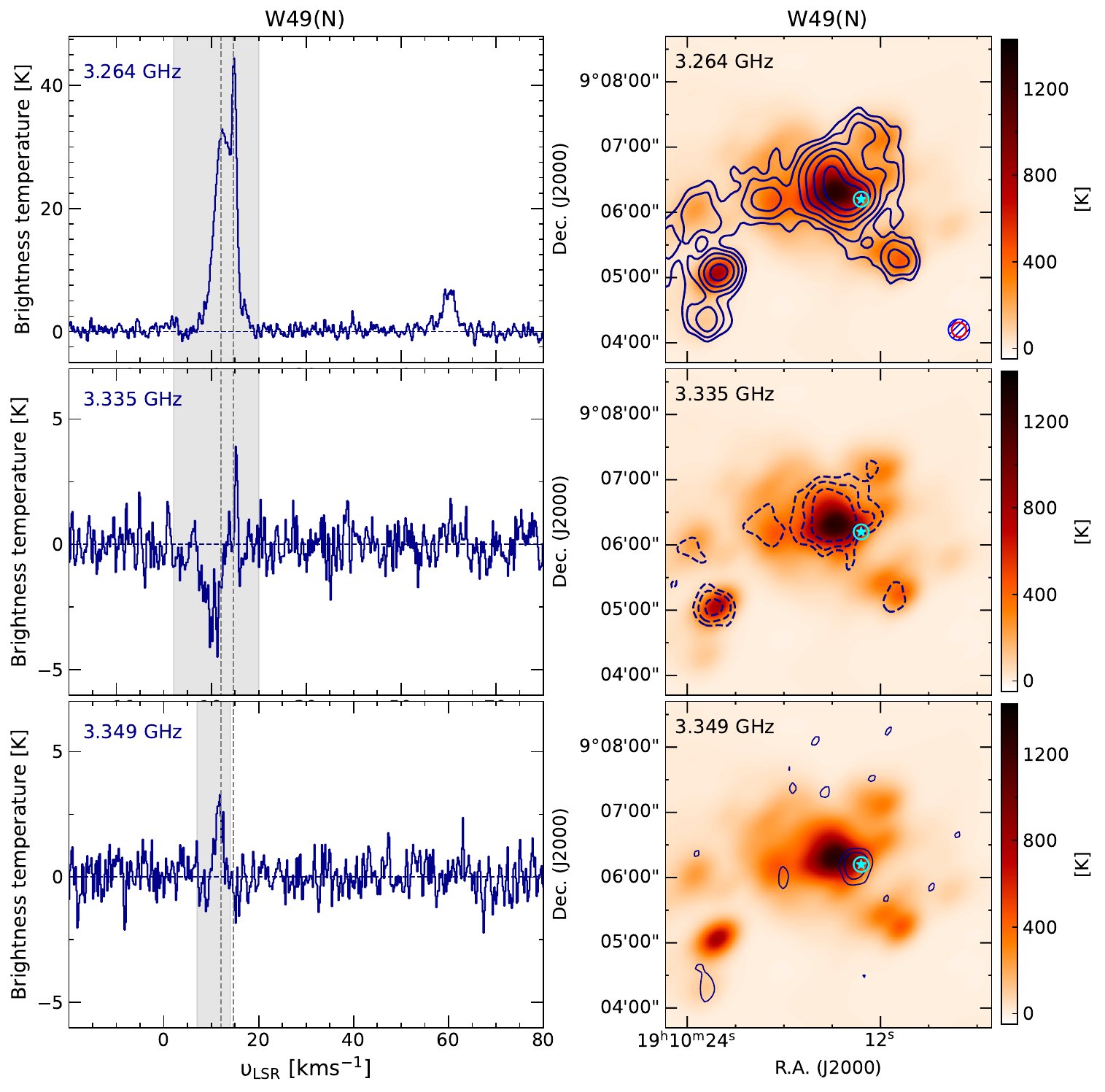}
    \caption{Left panel: (Top-to-bottom) Spectra of the ground state HFS splitting transitions of CH at 3.264~GHz, 3.335~GHz, and 3.349~GHz (in dark blue) towards W49~(N), extracted from the region in the map indicated by an encircled star (in cyan) presented on the right-hand panel. Dashed grey vertical lines mark the velocity components modelled. Right panel: Displays the integrated emission (solid) and absorption (dashed) intensity contours of the 3.264~GHz, 3.335~GHz, and 3.349~GHz (from top-to-bottom) lines, atop the 3.3~GHz continuum emission on brightness temperature scales. The velocity interval over which the features were integrated are highlighted in grey on the left panel. The beam sizes of the JVLA (blue and white filled, hatched circle), and the SOFIA/upGREAT (red) are displayed in the bottom right hand corner of the top-right panel, for comparison. This figure is produced based on data presented in \citet{Jacob2021}.}
    \label{fig:CH_maser-spectra}
\end{figure}

\begin{figure}
    \centering
    \includegraphics[width=0.65\textwidth]{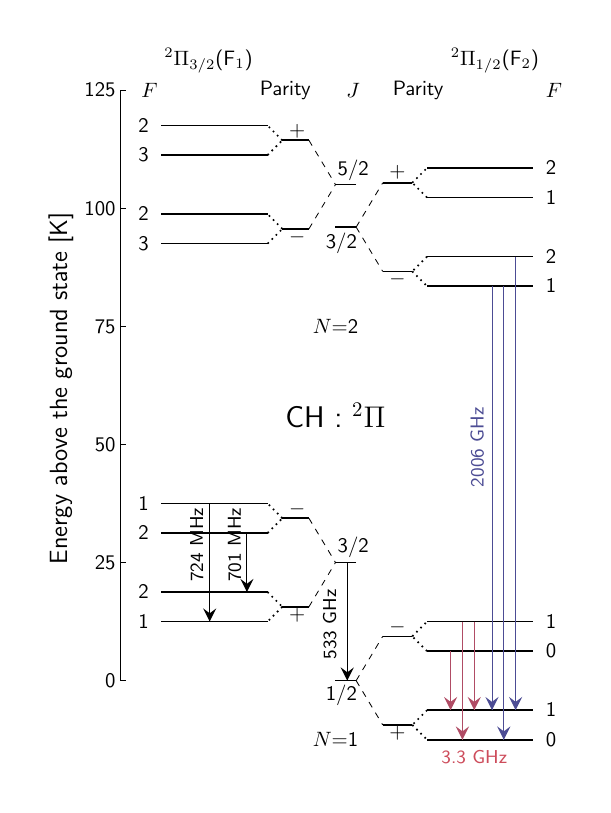}
    \caption{Lowest rotational energy levels of CH with the relevant transitions marked using arrows. Note that the separation of the $\Lambda$-doublet and HFS splitting levels are not drawn to scale. Adapted from \citet{Jacob2021}.}
    \label{fig:energy_level_diagram}
\end{figure}

In contrast to the sub-mm and FIR rotational transitions of CH (discussed in the previous section), accessible only from air-, and space-borne telescopes, the radio transitions of CH at 3.3~GHz have been extensively observed and with relative ease. The HFS splitting components corresponding to this transition are observed in (generally weak) emission toward a variety of environments ranging from quiescent dark clouds to H{\small II} regions, displaying line intensities that are inconsistent with the assumptions of LTE, where $I_{3.264~{\rm GHz}}:I_{3.335~{\rm GHz}}:I_{3.349~{\rm GHz}}=1:2:1$. As an example Fig.~\ref{fig:CH_maser-spectra} displays the spectra of the ground state HFS splitting transitions of CH toward W49~(N). The 3.264~GHz (lower satellite) line is seen in emission, with a line intensity enhanced by a factor of up to 20 with respect to its value expected at LTE, while the 3.335~GHz (main line) and 3.349~GHz (upper satellite) lines show a combination of (weak) emission and absorption. The ubiquity of the observed level inversion in the ground state lines of CH toward varied sources suggests the presence of a general pumping scheme that preferentially populates the upper level of the ground state $\Lambda$-doublet independent of the physical conditions probed. \citet{Bertojo1976} and \citet{Elitzur1977} initially proposed that this weak masing action in the CH ground state was caused by excitation through collisions (with atomic and molecular hydrogen) to the first rotational level (see the energy level schematic of CH shown in Fig~\ref{fig:energy_level_diagram}). However, \citet{Bujarrabal1984} soon demonstrated that collisions alone could not sufficiently account for the observed enhancements or anomalous excitation. Consequently, the inclusion of radiative processes becomes necessary to explain the observed line strengths of the CH ground state transitions in star forming regions. \citet{Zuckerman1975} and \citet{Bujarrabal1984} discussed the influence of FIR line overlap caused by either bulk velocity gradients in the gas or by thermal line broadening which bring the FIR rotational lines of CH in resonance with the ground state. The resulting partial or total line overlap can alter the degree of radiative trapping between the two ground state $\Lambda$-doublet levels and dictate their relative populations. Previous attempts to characterise the excitation of these lines by statistical equilibrium calculations were limited by the lack of reliable collisional rate coefficients and general uncertainties regarding the abundance of CH.

\begin{figure}
    \centering
    \includegraphics[width=1\textwidth]{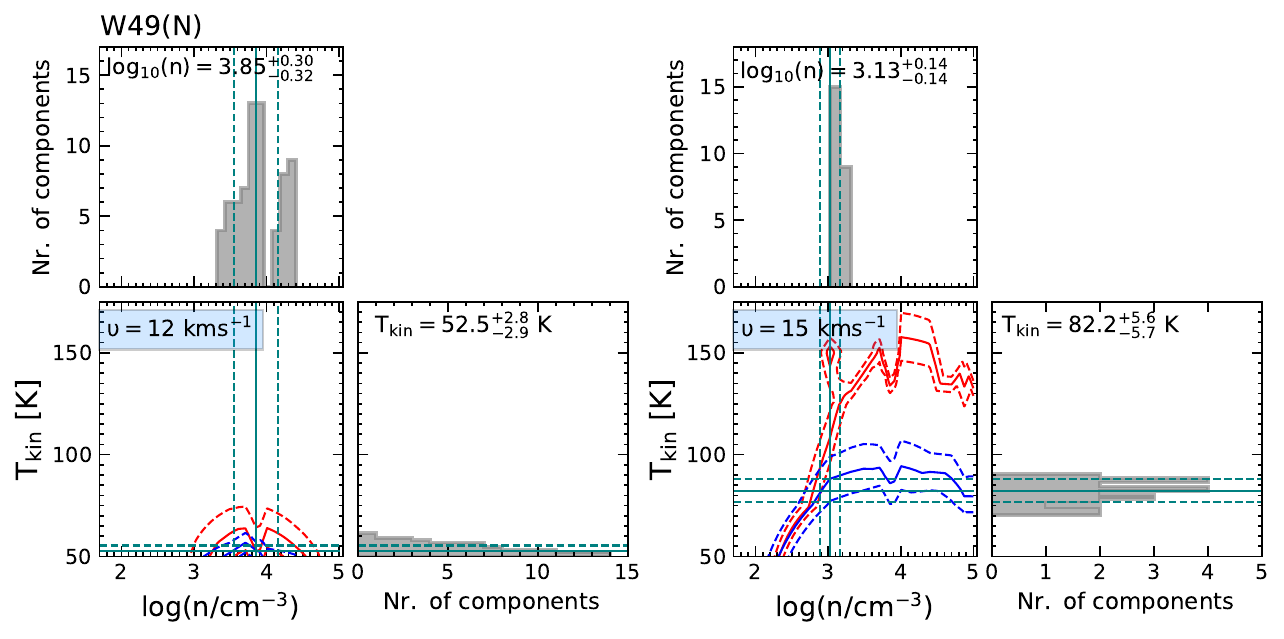}\\
    \includegraphics[width=0.46\textwidth]{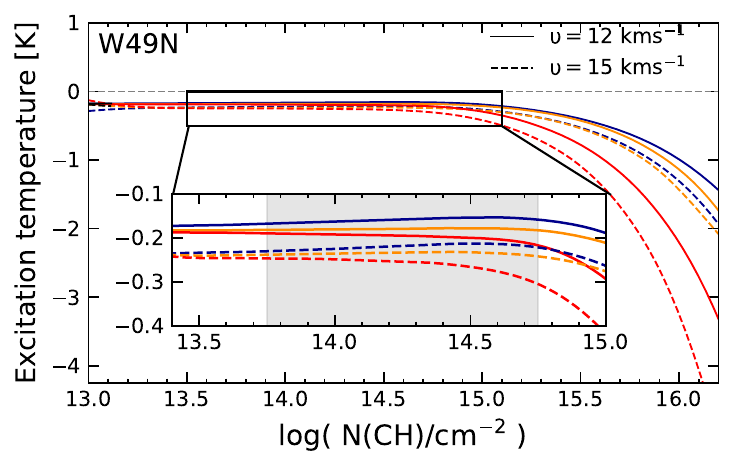}\quad
        \includegraphics[width=0.46\textwidth]{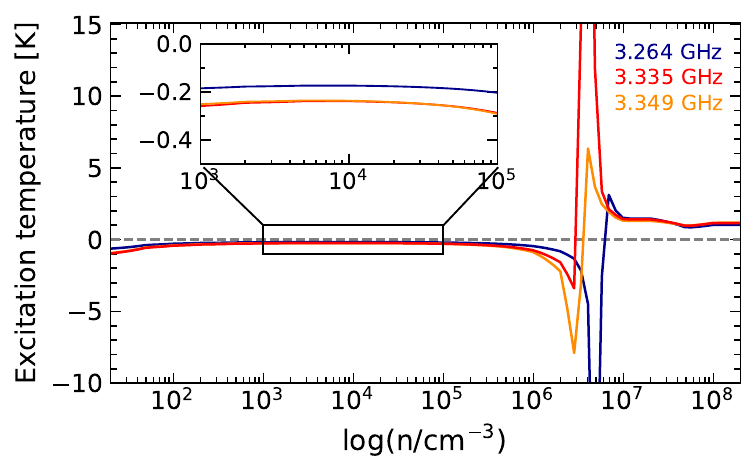}
    \caption{Top: Corner plot displaying the modelled gas density, and temperature distributions toward different velocity components in W49~(N). The solid and dashed red and blue curves represent the models that best reproduce the $T_{\rm 3.264~GHz}/T_{\rm 3.349~GHz}$ and $T_{\rm 3.264~GHz}/T_{\rm 3.349~GHz}$ line ratios and their errors. The parameter space is constrained by 3~$\sigma$ level of the minimum $\chi^2$ with the median and standard deviations marked by the solid and dashed teal lines, respectively. Bottom: The modelled excitation temperatures as a function of column densities (left) and gas densities (right) for a fixed set of physical conditions based on model constraints. The inset expands to show the excitation temperatures for the relevant range of CH column densities and gas densities. Reproduced with permission from \citet{Jacob2019}.}
    \label{fig:CH_excitation}
\end{figure}
Benefiting from recently computed collisional rate coefficients for inelastic collisions between CH and atomic/molecular hydrogen, as well as helium, by \citet{Dagdigian2018} and \citet{Marinakis2019}, \citet{Jacob2021} re-examined the issue of anomalous excitation in the CH ground state lines. \citet{Jacob2021} presented the first interferometric observations of CH using the NRAO\footnote{The National Radio Astronomy Observatory (NRAO) is operated by Associated Universities Inc., under a collaborative agreement with the US National Science Foundation.} \emph{Karl G. Jansky} Very Large Array (JVLA) in New Mexico. As an example, Figure~\ref{fig:CH_maser-spectra} presents an overview of the CH observations carried out toward W49~(N). These authors used the non-LTE radiative transfer code MOLPOP-CEP \citep{Asensio2018} which employs a coupled escape probability formalism to solve the radiative transfer equation for multi-level systems. Moreover, this code is capable of accounting for the effects of FIR line overlap within a plane-parallel slab geometry. The models were computed iteratively over a density-temperature grid of size $100\times100$, for $n_{\rm H}$ values in the range of 25~cm$^{-3}$ to 10$^5$~cm$^{-3}$ and gas temperatures between 50~K and 175~K. Contributions from the different collisional partners were weighted by the molecular fraction, ${f^{N}_{\text{H}_{2}} = 2N(\text{H}_{2})/\left( N(\text{H}) + 2N(\text{H}_{2}) \right)}$ and by assuming an ortho-to-para for H$_2$ ratio of 3:1. The models were constrained by using the column densities of CH determined from the $N, J=2,3/2\rightarrow1,1/2$ transitions of CH at 2~THz discussed in Sect.~\ref{subsec:CH_CO-darkgas}, which as shown in Fig.~\ref{fig:energy_level_diagram} shares a common lower energy level with the CH ground state. The physical conditions modelled, reveal that CH resides within warm cloud layers of photodissociation regions (PDRs), tracing gas layers with an average temperature of 70~K, with specific modelled components showing values as high at 125~K. This hints at the plausible formation of CH via endothermic reaction pathways involving CH$^+$, 
\begin{equation}
    {\rm C}^{+}+{\rm H}_2 \xrightarrow[\text{\cite{Hierl1997}}]{\text{4640~{\rm K}}} {\rm CH}^+ +{\rm H}_2;  \rightarrow {\rm CH}_2^+ + e^- \rightarrow {\rm CH}\, ,
    \label{eqn:CH_formation_pathway}
\end{equation}
initiated by shocks or the dissipation of turbulence or via the exchange of matter between the cold and warm neutral phases of the ISM \citep[for more details see][]{Godard2023}. 
Physically, such a chemical pathway driven by elevated temperatures may also be responsible for large scale velocity gradients in the gas which can result in line overlap between different sets of HFS splitting levels. The models that best reproduced the observed level inversion in the CH ground state lines yielded excitation temperatures $-0.3~$K, which fortuitously is consistent with early predictions by \citet{Bertojo1976}. Figure~\ref{fig:CH_excitation} displays the modelled excitation temperatures of the ground state transitions of CH as a function of CH column densities for the physical conditions derived toward different cloud components in W49~(N). Examining the modelled excitation temperatures as a function of gas densities, one would expect the ground state lines to be easily thermalised given their low Einstein A-coefficients (see Table~\ref{tab:CH-spec_properties}). Once the gas densities surpass the critical density (a few $\times10^{6}$~cm$^{-3}$) corresponding to the $N,J=1,3/2\rightarrow1,1/2$ rotational transition of CH near 532/537~GHz, the masing action quenches. Consequently, the excitation temperature increases, as the lines tend toward thermalisation. This further emphasises the importance of including the effects of FIR line overlap in understanding the masing present in the ground state CH lines. However, for higher modelled gas densities and a fixed gas temperature of 75~K, the CH lines do not approach thermalisation. This discrepancy may arise from the fact that the models presented by these authors did not include the effects of collisional excitation by electrons. \citet{Bouloy1984} had previously demonstrated that while collisions with electrons did not significantly produce level inversion in the CH ground state, it was responsible for thermalising these lines. Therefore, a comprehensive model that accurately describes the excitation of the ground state lines of CH is still pending the inclusion of collisional excitation by electrons, rate coefficients for which are presently not available. 

In another recent study \citet{Dailey2020} investigated the ground state excitation conditions of CH by comparing emission in the 3.335~GHz line with existing CH optical absorption from the 4300.2~\AA\ band toward a sample of 16 nearby stars. These authors noted sightline depended variations in the 3.335 GHz line’s excitation temperature and emphasised that the assumption that excitation temperature is greater than the background radiation temperature is not always true. The constraints hence placed on the excitation conditions of the ground state $\Lambda$-doublet lines of CH, restores the use of these lines as powerful probes of the diffuse and translucent ISM at radio wavelengths. Furthermore, these modelling efforts have motivated systematic efforts to extend observations of CH across the Milky Way (see discussion in Sect.~\ref{subsec:radio_wavelengths}) and in external galaxies.

\section{The diagnostic power of CH isotopologues}\label{subsec:CH_isotopes}
One of the many astrophysical applications of molecular spectroscopy is the determination of isotopic abundance ratios. This is because, measurements of the abundances of interstellar molecules and their isotopes when combined with theoretical models provide unique probes of the nucleosynthesis history of the region. Specifically, molecules and not atoms have been used for this purpose because their spectra better reflect the influence that the presence or absence of a neutron in the nucleus has on the electronic configuration of that species (i.e., isotopologues). In addition, molecules provide a rich chemistry of species from which isotopic abundance ratios can be determined.

However, isotopic ratios determined using interstellar molecules significantly deviate from their elemental abundances due to isotope-selective effects which include those due to optical depth effects, gas-phase chemical fractionation and selective photodissociation. For example, molecules susceptible to chemical fractionation -- caused by the small differences in the zero point energies between the two isotopes -- enhances the production of the rarer isotope, while photodissociation selectively destroys it because of its weaker self-shielding. In addition to these effects, other chemical processes must be involved in enhancing the observed isotopic abundance ratios of different molecular species that depend on the molecules’ environment. For this reason, the physical and chemical conditions that prevail in the regions where the molecules are formed also play an important role in enhancing the observed isotopic abundance ratios, making the ratio in turn a strong indicator of both the chemistry as well as the physical conditions probed.

\textit{Despite being the first molecule to be detected in space and subsequently across a wide range of frequencies, very little was known about the isotopologues of CH, -- $^{13}$CH and CD.} The following sections detail the detection of the CH isotopes and illustrate their diagnostic properties. 
These astronomical detections were facilitated by laboratory measurements of the rotational spectra of $^{13}$CH and CD made using techniques of high resolution laser magnetic resonance (LMR) spectroscopy \citep{Brown1989, Davidson2004}. Both $^{13}$CH and CD, like CH have a $X^{2}\Pi_{\rm r}$ ground state that conforms to a Hund’s case (b) coupling, with each total angular momentum level, \textbf{$J$}, splitting into two parity states (+ and $-$) due to $\Lambda$-doubling. In
addition, owing to the non-zero nuclear spins, $I$, of carbon-13 ($I$($^{13}$C) = 0.5) and deuterium ($I$(D) = 1), each rotational level further splits into HFS splitting levels, $F$ (= $J + I$). The HFS levels corresponding to the fundamental transitions of $^{13}$CH and CD were identified in the laboratory by \citet{Davidson2004} and \citet{Halfen2008}, respectively, using refined spectroscopic parameters \citep{Steimle1986, Wienkoop2003}, following which the rest frequencies of these transitions were also determined, up to an accuracy of a few 100 kHz. The frequencies of the transitions discussed in the following sections are summarised in Table.~\ref{tab:CH-isotopes-spec_properties}.

\subsection{\texorpdfstring{$^{13}$CH}{13CH}}\label{subsec:13CH}
The ubiquity of molecules containing carbon in the Universe has turned the ratio between the abundances of it’s two stable isotopes, $^{12}$C and $^{13}$C, into a cornerstone for Galactic chemical evolution studies. This is because unlike $^{12}$C which is formed as the primary product of He burning in massive stars, $^{13}$C is formed as a secondary product of stellar nucleosynthesis, making its ratio a measure of the primary to secondary processing or the nuclear history of a Galaxy. Early chemical evolution models \citep[see for example,][]{Guesten1982} predicted that this ratio would decrease as a result of an enrichment in the secondary products over galactic timescales. Observationally, this ratio displays an overall rising gradient with galactocentric radii, when measured using observations of different carbon-bearing molecules like CO (C$^{18}$O) \citep{Langer1990, Wouterloot1996, Giannetti2014}, CN \citep{Savage2002, Milam2005}, H$_2$CO \citep{Henkel1985, Yan2019} and even complex organic molecules \citep{Halfen2017}. However, this gradient shows systematic variations depending on the species, which arises from isotope-selective effects (discussed above) that affect different molecules, differently. This includes optical depth effects, gas-phase chemical fractionation and selective photodissociation. In addition, the high optical depths of many abundant $^{12}$C-bearing molecules are affected by saturation and self-absorption effects which in turn skew estimates of the subsequently derived $^{12}$C/$^{13}$C ratio.

CH forms a potentially unbiased tool for measuring the $^{12}$C/$^{13}$C ratio since it is free from optical depth effects (as discussed in Sect.~\ref{subsec:CH_CO-darkgas}) and formed at an early-stage of ion-molecule chemistry via endothermic chemical pathways involving C$^+$, and driven by elevated temperatures (see Eq.~\ref{eqn:CH_formation_pathway}). For this reason, CH is relatively unaffected by the effects of fractionation as in the case of CH$^+$ \citep{Ritchey2011} and if anything reflects the degree of fractionation in C$^+$. The effects of fractionation in carbon-bearing species has been examined in great detail by \citet{Rollig2013} using detailed PDR models. Their results also point to the fractionation present in CH being dominated by the fractionation of its parental species, C$^+$. \\

\begin{figure}
    \centering
    \includegraphics[width=0.6\textwidth]{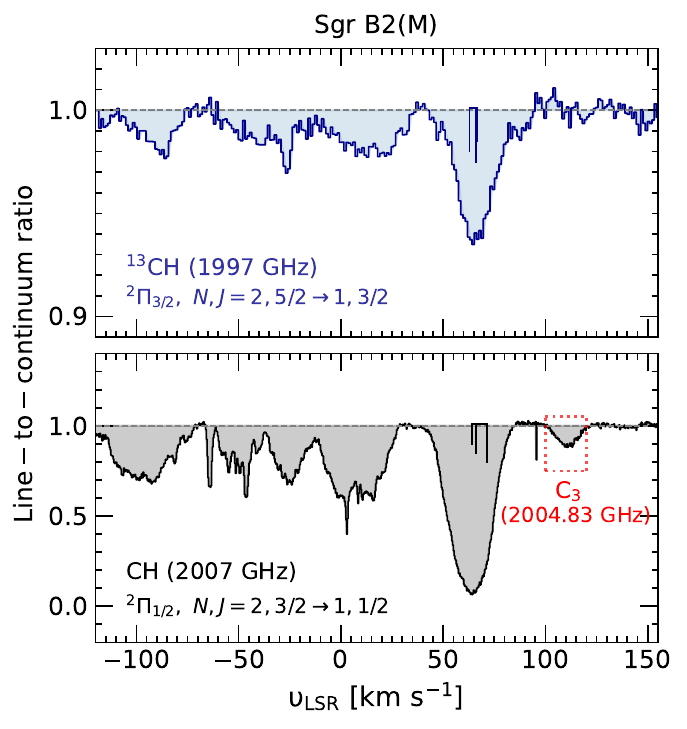}
    \caption{Main-beam continuum normalised spectra of $N, J = 2, 3/2 \rightarrow 1, 1/2$ transition of $^{13}$CH near 1997~GHz (top) and $^{12}$CH near 2007~GHz (bottom) toward Sgr~B2(M) in blue and black, respectively. The $^{12}$CH spectrum is contaminated by C$_{3}$ absorption at 2004.833~GHz arising from the image band, as indicated by the dotted red box. This figure is created based on data presented in \citet{Jacob202013CH}.}
    \label{fig:12CH-13CH-sgrb2m}
\end{figure}

While, \citet{Richter1967} reported the identification of $^{13}$CH in the solar spectrum, early attempts to detect $^{13}$CH in the general ISM were fruitless. For example, \cite{Bottinelli2014} searched for the $N,J = 1,1/2 \rightarrow 1,3/2$ and $N,J = 1,3/2 \rightarrow 2,5/2$ transitions of $^{13}$CH toward the well studied low-mass protostar IRAS~16293$-$2422 using Herschel/HIFI but were unsuccessful. However, with the end of the Herschel mission, only SOFIA was capable of providing an avenue for procuring high resolution data of the fundamental rotational transitions of CH. In addition the transitions of CH observable using SOFIA/upGREAT have the advantage of being observed in deep absorption, promising the measurement of reliable abundances unlike in the case of the 532/536~GHz lines observed using Herschel/HIFI, that display complex line profiles with a mixture of emission and absorption. Therefore, \citet{Jacob202013CH} searched for the HFS splitting lines corresponding to the $N, J=2,3/2\rightarrow 1, 1/2$ 2~THz transitions of $^{13}$CH near 2~THz. These authors reported the first detection of $^{13}$CH in the ISM toward four star-forming regions, Sgr~B2(M), G34.26+0.15, W49~(N) and W51~E and a non-detection toward W3(OH). Figure~\ref{fig:12CH-13CH-sgrb2m} displays the $^{13}$CH spectrum observed toward Sgr~B2(M). Similar to the $^{12}$CH absorption line profile, the $^{13}$CH spectrum shows deep absorption not only at the systemic velocity of the background molecular cloud near 64~km~s$^{-1}$ but also at blue shifted velocities along its line-of-sight. Note that in the specific example presented here, of the CH 2~THz transition toward Sgr~B2(M), the CH spectrum is contaminated by an absorption feature at 2004.833~GHz \citep{Gendriesch2003} arising from the other sideband of the upGREAT receiver and identified as being due to high-lying transitions of C$_3$. \\

\begin{figure}
    \centering
    \includegraphics[width=0.65\textwidth]{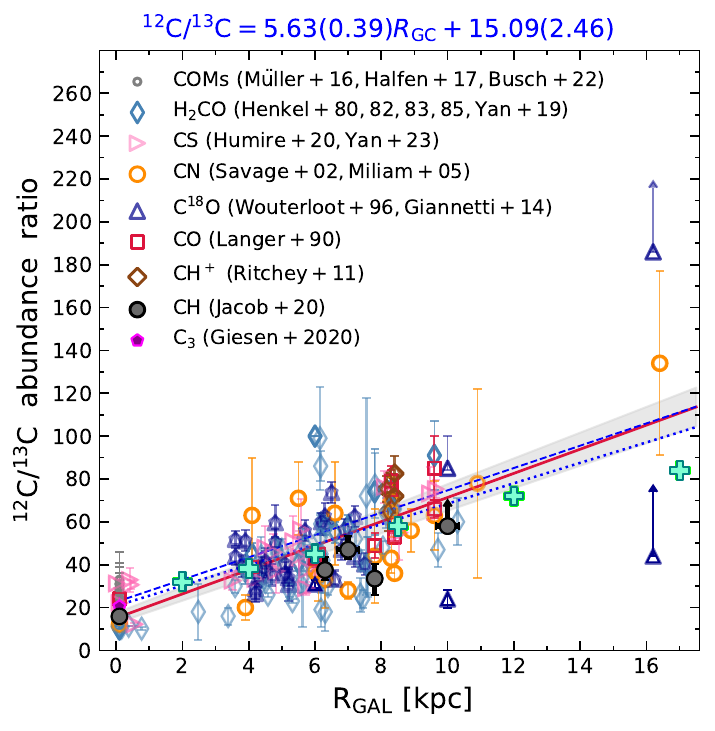}
    \caption{Plot of $^{12}$C/$^{13}$C isotope ratios as a function of galactocentric distance, $R_{\rm GC}$ (kpc). The filled black circles represent the $^{12}$C/$^{13}$C ratio derived using CH \citep{Jacob202013CH}, while the unfilled grey points, narrow, blue diamonds, pink right facing triangles, orange circles, dark-blue upward facing triangles, red squares, brown diamonds and filled magenta pentagons represent values derived from complex organic molecules \citep[COMS or molecules containing six or more atoms;][]{Muller2016, Halfen2017, Busch2022}, H$_2$CO \citep{Henkel1980, Henkel1982, Henkel1983, Henkel1985, Yan2019}, CS \citep{Humire2020, Yan2023}, CN \citep{Milam2005}, C$^{18}$O \citep{Wouterloot1996}, CO \citep{Langer1990}, CH$^+$ \citep{Ritchey2011} and C$_3$ \citep{Giesen2020}, respectively. The red solid line represents the weighted fit to the data such that, $^{12}{\rm C}/^{13}{\rm C} = 5.63(0.39)R_{\rm GC} + 15.09(2.46)$, and the grey shaded region demarcates the 2~$\sigma$ interval of this fit. For comparison, the fit obtained by \citet{Halfen2017} and \citet{Yan2023} are displayed by the dashed, and dotted blue lines, respectively. In addition, the teal crosses plot the results of Galactic chemical evolution models discussed in \citet{Kobayashi2011} (see Sect. 4.9 of \citet{Yan2023} for a detailed discussion).}
    \label{fig:12C-13C-gradient}
\end{figure}

Combining the $^{12}$C/$^{13}$C ratio derived from CH with those of previous measurements made using other carbon-bearing species, we present here a revised Galactic gradient with overall reduced errors derived, such that $^{12}$C/$^{13}$C = 5.63(0.39)$R_{\rm GC}$ + 15.09(2.46) (see Fig~\ref{fig:12C-13C-gradient}). Not only does the $^{12}$CH/$^{13}$CH ratio potentially reflect the underlying $^{12}$C/$^{13}$C isotopic abundance, providing new constraints to Galactic chemical evolution models but observations of $^{13}$CH also aids in our understanding of $^{13}$C substitution and fractionation in subsequently formed species. The derived gradient is also consistent with the $^{12}$C/$^{13}$C abundance gradient predicted by the Galactic chemical evolution models presented in \citet{Kobayashi2011} with deviations only at larger Galactocentric radii ($>8$~kpc). This is because the trend is observationally unconstrained in the outer Galaxy, necessitating the need for robust measurements of the $^{12}$C/$^{13}$C ratio at these distances.

Furthermore, constraints on the $^{12}$C/$^{13}$C ratio are necessary to accurately interpret the isotopic abundance ratios of other species such as that of nitrogen ($^{14}$N/$^{15}$N), since this ratio is frequently derived from double isotopes, like the H$^{12}$C$^{14}$N, H$^{13}$C$^{15}$N pair \citep{Colzi2020}. Similarly, knowledge of the $^{12}$C/$^{13}$C ratio is essential for analysing the complex line profiles of C$^+$ (an important molecular gas tracer and coolant in the ISM, see discussion in Sect.~\ref{subsec:CH_CO-darkgas}) and estimating its optical depth. Comparisons between the optically thin $^{13}$C$^+$ line profiles and the observed $^{12}$C$^+$ profiles are used to distinguish self-absorption features from multiple cloud components, as demonstrated in \citet{Guevara2020}. As to external galaxies, $^{13}$CH has recently been detected in the southwestern component of the $z=0.89$ absorber towards the lens magnified quasar PKS~1830$-$211 by \citet{Muller2023}. These authors reported the detection of the $^{13}$CH lines corresponding to the $\Lambda$-doublet levels of the $N=1, J=3/2 \rightarrow 1/2$ transitions near 532~GHz and 536~GHz (rest frequency) using ALMA. By simultaneously fitting the $\Lambda$-doublet components of both $^{12}$CH and $^{13}$CH, \citet{Muller2023} derive a $^{12}$CH/$^{13}$CH ratio of 150$\pm$10. As noted by these authors, the $^{12}$C/$^{13}$C ratio derived from CH is higher than that derived by other species including, 97$\pm$6 when using CH$^+$ \citep{Muller2017} and values as low as 20--50 from HCO$^+$, HCN and HNC \citep{Muller2011}. While, these differences can once again be attributed to isotope selective effects, the larger $^{12}$C/$^{13}$C abundance ratio derived using CH is consistent with what is expected in poorly processed environments such as the material probed along this line-of-sight \citep{Kobayashi2011}. This further supports the use of CH as a probe of the true $^{12}$C/$^{13}$C elemental isotopic ratio.

\subsection{CD}\label{subsec:CD}

Early observational searches for deuterium in the ISM were motivated by the constraints that measurements of its abundances provided on cosmology. This stems from the deuterium atom's unique chemical history and subsequent processing. Interstellar deuterium is formed in significant amounts at the birth of the Universe through primordial nucleosynthesis alongside other light elements (He, Li) but unlike these other species deuterium is always net destroyed as it rapidly gets processed in stellar interiors. While the most useful estimates of the D/H abundance ratio comes from direct measurements of D{\small I} and H{\small I} absorption lines in damped Lyman $\alpha$ (DLAs) systems at redshifts, $z = 2$ -- $3$ \citep{Khersonsky1995}, placing the primordial D/H ratio at $\sim 2.5\times10^{-5}$ \citep{Cooke2018}, such observations are limited to only a handful of detections. 

An alternative method for determining the D/H ratio is through observations of deuterated molecules and their hydrogenated counterparts. However, observations quickly revealed that the D/H abundance ratio derived from deuterated molecules, typically have abundances far greater than the elemental D/H abundance ratio, with values as high as 30--70~\% \citep[e.g., ND/NH toward IRAS 16293$-$2422,][]{Bacmann2010}. As in the case of the $^{12}$C/$^{13}$C ratio discussed above (Sect.~\ref{subsec:13CH}), these enhancements in the D/H ratio result from isotopic fractionation effects, which favours the production of the deuterated isotope over its hydrogenated counterpart. Today, over 50 deuterated molecules have been detected in space showing a range of deuterium substitutions from singly-deuterated species like HD \citep{Lacour2005} to triply-deuterated species like ND$_3$ \citep{Lis2002} and CD$_3$OH \citep{Parise2004}. Formed under very specific physical conditions, these highly-deuterated molecules instead of probing cosmology make for very interesting probes of their environments and chemistry \citep[see][for a review]{Ceccarelli2014}.

In molecular clouds, deuterium is primarily locked into HD which then initiates deuterium transfer through exothermic ion-molecular reactions involving H$_3^+$ in the cold ISM ($T \leq 20~$K) (reaction \ref{eqn:deu1}). At warmer temperatures ($T\sim 30$--80~K) deuterium fractionation is driven by ion-molecular reactions of HD with CH$_2$D$^+$ and C$_2$HD$^+$ (reactions \ref{eqn:deu2} and \ref{eqn:deu3}).
\begin{align}
    {\rm H}_3^+ + {\rm HD} & \rightarrow {\rm H}_2{\rm D}^+ + {\rm H}_2 \hspace{1.65cm} \Delta E = 230~{\rm K}; \text{\citealt{Gerlich2002}} \label{eqn:deu1}\\
    {\rm CH}_3^+ + {\rm HD} & \rightarrow {\rm CH}_2{\rm D}^+ + {\rm H}_2 \hspace{1.4cm} \Delta E = 390~{\rm K}; \text{\citealt{Asvany2004}} \label{eqn:deu2}\\
    {\rm C}_2{\rm H}_2^+ + {\rm HD} & \rightarrow {\rm C}_2{\rm HD}^+ + {\rm H}_2 \hspace{1.4cm} \Delta E = 550~{\rm K}; \text{\citealt{Herbst1987}} \label{eqn:deu3}
\end{align}

\begin{figure}
\hspace{-1cm}
    \includegraphics[width=1.2\textwidth]{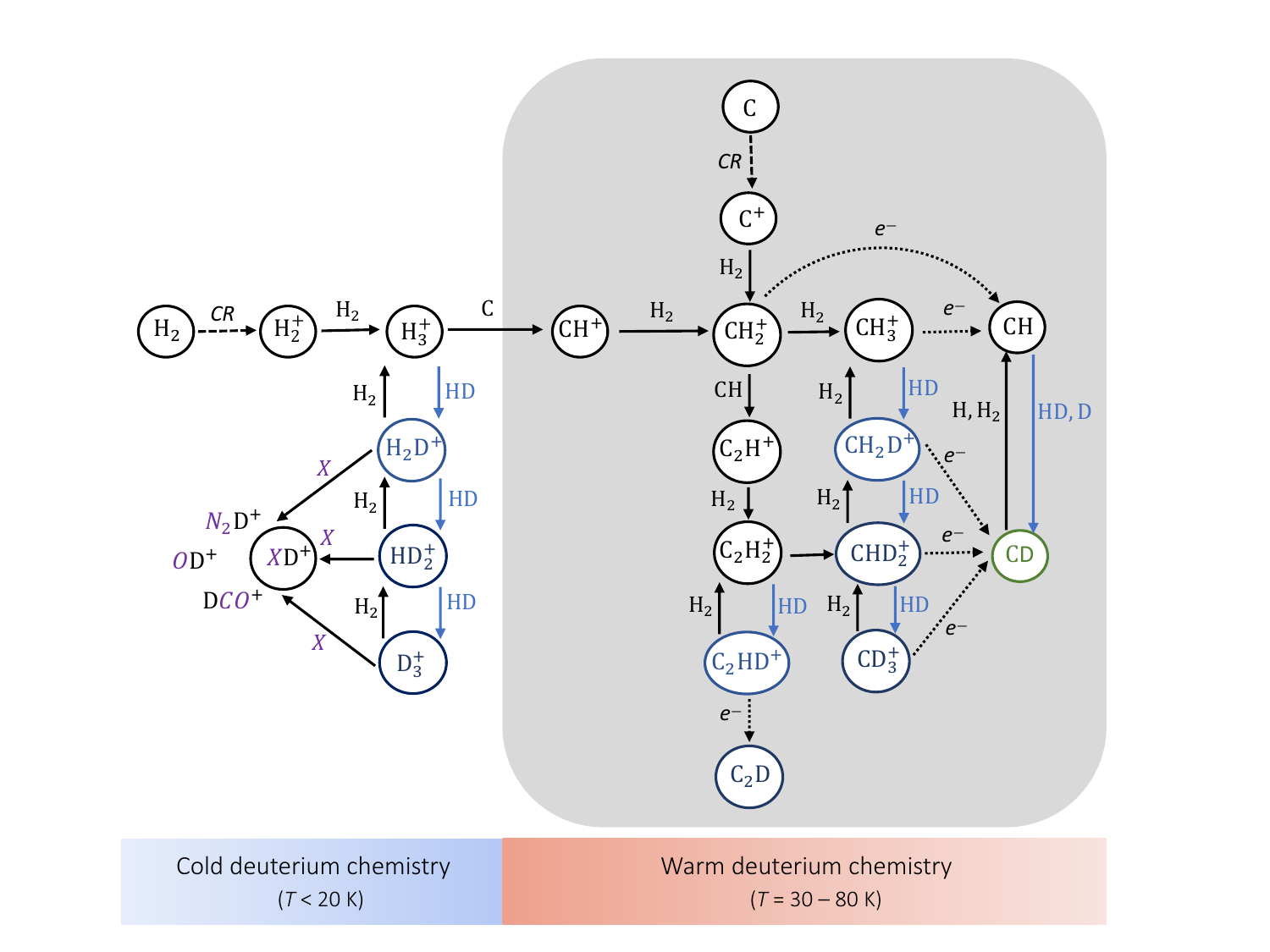}
    \caption{Network displaying the key ion-neutral and neutral-neutral reactions involved in deuterium fractionation.}
    \label{fig:deuterium_network}
\end{figure}
While proton-deuteron reactions in cold-cloud conditions involving H$_2$D$^+$ have been widely explored \citep{Caselli2003, Flower2004, Walmsley2004, Pagani2009, Brunken2014}, very little is known about deuteration at intermediate/warmer temperatures, i.e., temperatures in which contributions from CH$_2$D$^+$ and C$_2$HD$^+$ become significant relative to H$_2$D$^+$. Being a central species in interstellar carbon-chemistry, observations of CH$_3^+$ in particular, alongside that of its deuterated counterparts (CH$_2$D$^+$, CHD$_2^+$, CD$_3^+$) would present unique probes of warm deuteration as well as progress our understanding of the degree of deuteration inherited in subsequently formed species. For example, enhancements in the abundances of species like DCN in warm gas have been interpreted to result from high degrees of deuteration in its progenitor, CH$_2$D$^+$ \citep{Leurini2006}. 

Observationally, owing to its symmetric and planar structure CH$_3^{+}$ does not possess a permanent dipole moment and hence has no observable rotational transitions. Berne et al. 2023 have recently identified emission features in the JWST/MIRI spectrum towards a protoplanetary disk, d203$-$506, in the Orion Bar\footnote{Observations carried out under the umbrella of the JWST ERS program PDRS4all \citep{Berne2022}.}, which are coincident with the vibrational bands of CH$_3^+$ near 7~$\mu$m. While these observations are facilitated by laboratory measurements of the vibrational lines \citep{Asvany2018}, as noted by these authors additional high-resolution laboratory measurements of the infrared spectrum of CH$_3^+$ near 7~$\mu$m are necessary for identifying the individual features observed in the spectra. As to deuteration, the deuterated counterparts of CH$_3^+$ have rotational transitions which are observable at mm-, and radio wavelengths. But to date, only tentative detections of two transitions of CH$_2$D$^+$ have been reported toward Orion~IRc2 by \citet{Roueff2013}. For this reason, observers turned their attention toward early by-products in the deuterated-carbon network, such as CD, formed via the dissociative recombination of CH$_2$D$^+$ (CHD$_2^+$, CD$_3^+$). 

\begin{figure}
    \centering
    \includegraphics[width=1\textwidth]{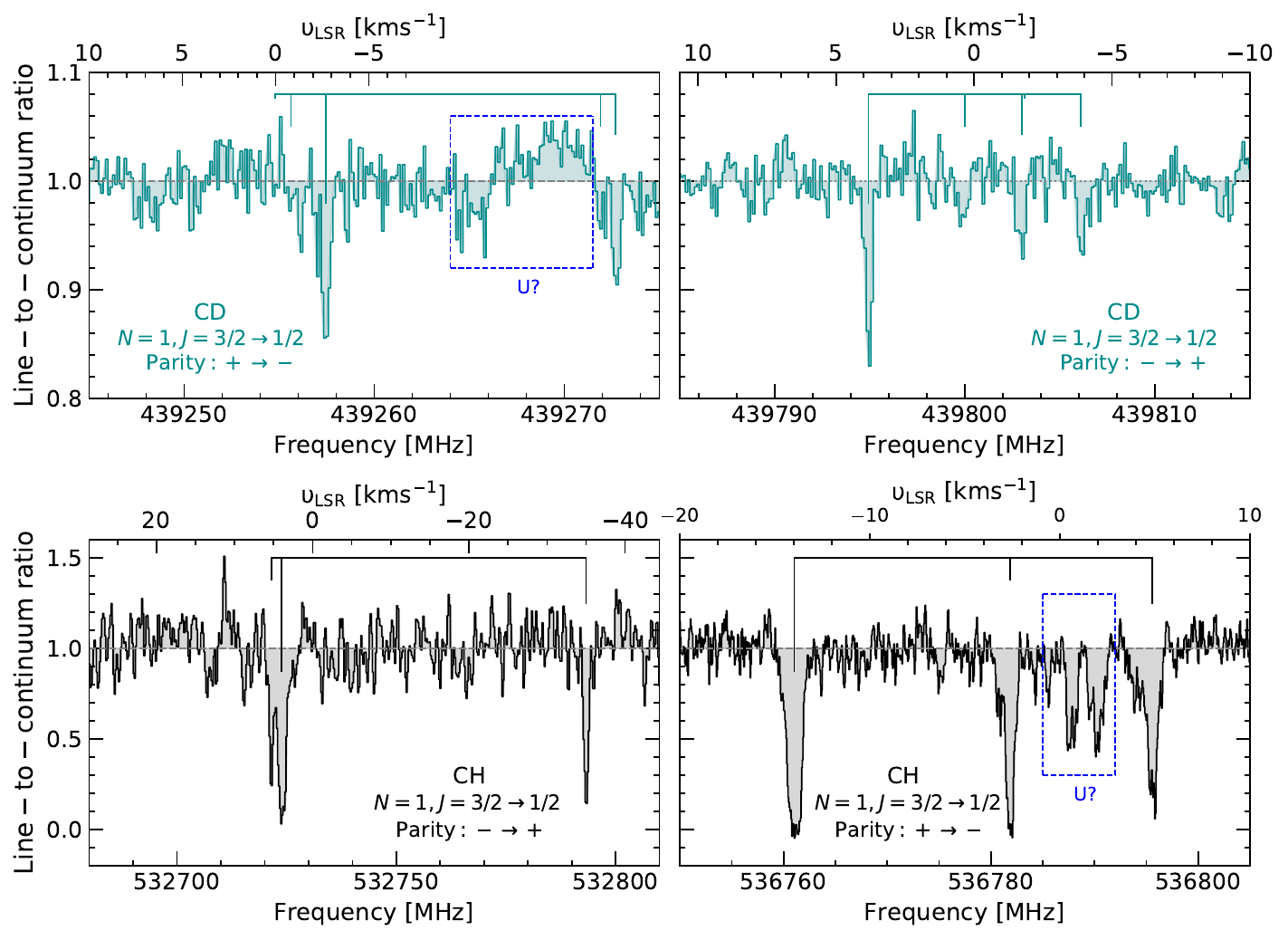}
    \caption{Spectra of the $N=1, J = 3/2\rightarrow 1/2$ transitions of CD (top-panel, in teal) and CH (bottom-panel, in black). Unidentified, contaminating spectroscopic and/or baseline features are labelled- U? in blue. This figure is created based on data presented in \citet{Jacob2023} and \citet{Bottinelli2014}.}
    \label{fig:CD_spec}
\end{figure}

\begin{figure}
    \includegraphics[width=\textwidth]{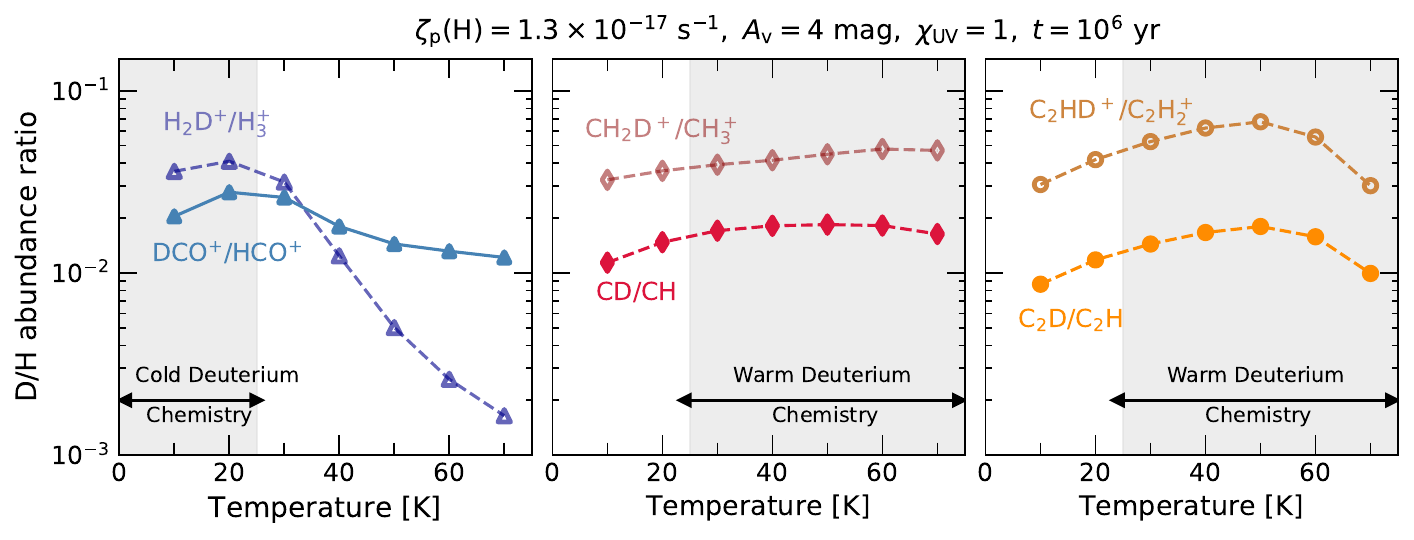}
    \caption{Left-to-right: Modelled D/H abundance ratios as a function of gas temperatures, from left to right for HCO$^+$ (filled blue triangles), CH (filled red diamonds) and C$_2$H (filled yellow circles) alongside the D/H ratios obtained from their progenitor species H$_3^+$ (unfilled light blue triangles), CH$_3^+$ (unfilled light red diamonds) and C$_2$H$_2^+$ (unfilled dark yellow circles) at a simulation time, $t =10^{6}~$years \citep{Brunken2014}. }
    \label{fig:temp_dependence}
\end{figure}

\citet{Jacob2023} reported the detection of the $^{2}\Pi_{3/2}, N=1, J=3/2 \rightarrow ^{2}\Pi_{1/2}, N=1, J=1/2$ HFS splitting transitions of CD near 440~GHz (see Table~\ref{tab:CH-isotopes-spec_properties}) toward IRAS~16293$-$2422\footnote{IRAS~16293$-$2422 is a low-mass protostar located within the Ophiuchus star-forming complex and hosts a rich chemistry, owing to which it has been extensively studied, for example under the ALMA Protostellar interferometric line survey \citep[PILS;][]{Coutens2016}.}, using the APEX/nFLASH460 receiver. Figure~\ref{fig:CD_spec} presents the observed $\Lambda$-doublet lines of CD alongside the corresponding CH lines near 532/536~GHz presented in \citet{Bottinelli2014}. The derived D/H abundance ratio of (0.016$\pm$0.23) determined from CH (CD) is comparable to that of HCO$^+$ and HCN while it is an order of magnitude lower than that derived from C$_2$H and H$_2$CO \citep[see comparison presented in Table.~3 of][]{Jacob2023}. Utilising simulations run under the framework of the gas-grain chemical code, pyRate \citep{Sipila2015, Sipila2019}, for a fixed set of parameters \citep[see Sect.~5.1 of][]{Jacob2023}, these authors first confirmed that the dissociative recombination of CH$_2$D$^+$ with electrons was the dominant formation pathway for CD under conditions that typically represent envelope cloud conditions of IRAS~16293$-$2422. Following which they investigated the temperature dependence of the CD/CH ratio (Fig.~\ref{fig:temp_dependence}). The D/H ratio traced by CH peaks at gas temperatures $\sim\!40$--60K. However, beyond this temperature range, the D/H ratio decreases, similar to its parent species CH$_2$D$^+$ and consistent with predictions from previous theoretical models \citep{Millar2005}. The temperature dependence of deuteration in early products of interstellar chemistry, such as HCO$^+$, CH, and C$_2$D$^+$, shown in Fig.~\ref{fig:temp_dependence}, closely resembles that of their respective precursor species, namely, H$_2$D$^+$, CH$_2$D$^+$, and C$_2$HD$^+$. Therefore, the successful detection of CD toward IRAS~16293$-$2422 provides new evidence for chemistry initiated by reactions involving CH$_2$D$^+$, placing constraints on the temperature of the outer envelope of this protostar. Dedicated searches for CD toward other regions are required to better our understanding of the proton-deutron transfer in warm interstellar environments and the degree of deuteration in subsequently formed species.

Although deuterated molecules besides HD have been detected toward nearby galaxies \citep{Chin1996, Heikkila1997, Martin2006}, the $z=0.89$ absorber towards PKS~1830$-$211 promises to be an excellent target for extending the search for CD. This is not only because both CH and $^{13}$CH have been successful detected towards this system but also because of the detection of several deuterated species towards it (for e.g., upper limits for DCO$^+$, DCN \citep{Shah1999, Muller2006}, HDO, ND, NH$_2$D; \citep{Muller2020}). The successful detection of CD toward this system of absorbers will potentially provide evidence for deuterium chemistry taking place via warm deuteration pathways in extragalatic systems. These observations may also aid in the interpretation of the high, observed degrees of deuteration towards this system, which was otherwise thought to take place via cold deuteration pathways.

\section{Searching for the elusive methylene radical}\label{sec:elusive_ch2}
Nobel laureate Gerhard Herzberg, a pioneer in laboratory astrophysics, was awarded the Nobel prize in chemistry in 1971, for his contributions to the knowledge of the electronic structure and geometry of molecules, particularly that of free radicals. Amongst the many discoveries he had made throughout his illustrious career, Herzberg later recounted the rightful identification of CH$_2$ in the laboratory, as being one that gave him a great deal of satisfaction. A search that spanned nearly two decades, the successfully assignment of the observed absorption bands near 1414.5~\AA\ as arising due to CH$_2$ by Herzberg using flash-photolysis techniques, paved the way for investigations of the internal structure and energy states of molecules.

Despite its importance in astrochemistry, much like in the laboratory the nature of CH$_2$ remains elusive in space. The following sections detail the hunt for the elusive CH$_2$ in space and the efforts undertaken to arrive at a similarly satisfying explanation for the source of its observed emission.

\subsection{The mysterious \texorpdfstring{CH$_2$}{CH2} emission}\label{subsec:CH2_mystery}
The methylene radical, CH$_2$, like CH (discussed in previous sections) is formed by the dissociative recombination of the methyl ion (CH$_3^+$) with an electron and is destroyed by reactions with atomic oxygen in dense clouds to form HCO and HCO$^+$, progenitors for the formation of more complex interstellar molecules. Theoretically, chemical models by \citet{black1978models}, \citet{prasad1980model}, and \citet{van1986comprehensive} predicted large abundances of CH$_2$ in diffuse and translucent clouds similar to that of CH. However, despite cementing its importance in interstellar chemistry, there have only been a handful of detections of CH$_2$ in the ISM, unlike its ubiquitous chemical counterpart, CH. 

\begin{figure}
    \centering
    \includegraphics[width=0.6\textwidth]{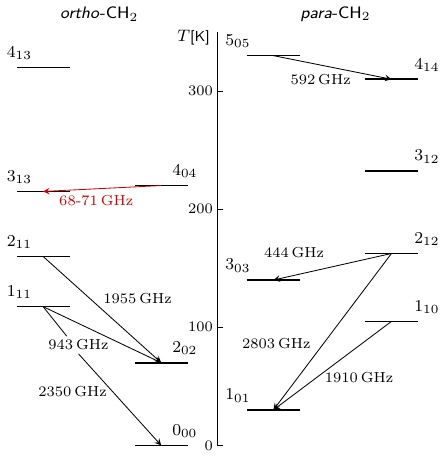}
    \caption{Ground state rotational energy level diagram of \textit{ortho}-, and \textit{para}-CH$_2$ up to an energy of 350~K. Highlighted in red is the transition discussed further in this work.}
    \label{fig:CH2_energy_level}
\end{figure}

This can be partly attributed to the energy level structure of CH$_2$ \citep{Lovas1983, Michael2003, Brunken2005}, which due to its peculiar b-type selection rules results in widely spaced energy levels with only four rotational transitions lying below 1000~GHz, between both its nuclear spin isomers, \textit{ortho}-, and \textit{para}-CH$_2$ (see Fig.~\ref{fig:CH2_energy_level}). Therefore, astronomical observations of both spin flavours of CH$_2$ are difficult as many of its transitions are inaccessible from the ground, while those observable from the ground lie close to the edges of atmospheric windows. As shown in Fig.~\ref{fig:CH2_energy_level}, the rotational energy levels of the $\boldsymbol{^{3}B_{1}}$ ground electronic state of CH$_2$ are characterised by the rigid-body angular momentum quantum number, $N$, and the pseudo-quantum numbers $K_{\rm a}$ and $K_{\rm c}$, which represent projections of $N$ along the prolate and oblate symmetry axes \citep{Herzberg1971}. Furthermore, having an electronic spin quantum number, $S$, of unity, the total angular momentum quantum number, $J$, splits into three levels with values equal to $N-1$, $N$, and $N+1$ for $N>0$. Spin-spin and spin-rotation interactions split the $N_{K_{\text{a}}K_{\text{c}}}$ rotational levels into three fine-structure levels. Furthermore, the fine-structure levels of \textit{ortho}-CH$_2$ which possesses a hydrogenic nuclear spin, $I_{\rm H}$, of unity are split by hyperfine-structure effects into energy levels characterised by the total angular momentum, $F = J-1$, $J$, and $J+1$ for $J\neq 0$, while those of \textit{para}-CH$_2$ with $I_{\rm H} = 0$, are not split further.

In spite of these observational difficulties, CH$_2$ has been detected in the ISM a handful of times, including for the first time via the $N_{KaKc}$ = $4_{04}-3_{13}$
lines of \textit{ortho}-CH$_2$ between $68-71$~GHz by \citet{Hollis1995}, tentatively at optical bands near 1397~\AA\ and 1410~\AA\ by \citet{Lyu2001} and \citet{Welty2020}, as well as at FIR wavelengths with definitive detections of both nuclear spin states of CH$_2$ at 107~$\mu$m and 127~$\mu$m by \citet{Polehampton2005}, respectively.

The original detection of the $N_{KaKc}$ = $4_{04}-3_{13}$ transitions of \textit{ortho}-CH$_2$ by \citet{Hollis1995} towards dense, and hot regions like Orion-KL and W51~Main contradicted chemical model predictions of the species in more diffuse environments. Nevertheless the detection of CH$_2$ toward these regions was unsurprising given that the observed transitions lie at an upper level energy of 225~K above the ground state. However, the association of the observed CH$_2$ emission in the Orion~KL region with the hot core, quickly came into question. This was because \citet{Hollis1995} carried out these observations using the NRAO Kitt Peak (KP) 12~m radio telescope in Arizona, which has a full width at half maximum (FWHM) beam width of 86$^{\prime\prime}$ at 70~GHz. Owing to the large beam size of the KP 12~m dish, it is likely that their observations covered several components associated with the Orion molecular cloud complex within a single beam. This suggested that the CH$_2$ emission may not arise from hot cores, a hypothesis that was further solidified by the non-detection of 68-71~GHz lines of CH$_2$ when observed towards the same position in Orion~KL but using instruments with smaller beam sizes at 70~GHz, like the 100~m Green Bank Telescope (GBT; FWHM = 12$^{\prime\prime}$; data presented in \citet{Frayer2015}) and the Institut de Radioastronomie Millim\'etrique (IRAM) 30~m telescope (FWHM = 35$^{\prime\prime}$). If the CH$_2$ emission did arise from the Orion-KL hot core, then the emission expected to be observed when using other telescopes can be predicted by scaling the peak intensity obtained from the KP 12~m telescope by the inverse of the beam filling factor\footnote{The beam filling factor is given by $({\theta_{\rm S}^2 + \theta_{\rm B}^2)/\theta_{\rm S}^2}$, where $\theta_{\rm S}$ and $\theta_{\rm B}$ are the FWHM source, and beam sizes, respectively}. Assuming a source size of 12$^{\prime\prime}$ \citep{Pauls1983}, results in peak intensities that are 5.5 and 26 times higher on main-beam brightness temperature scales, for the IRAM 30~m telescope and the GBT observations, respectively, relative to the KP observations. While this confirms that the CH$_2$ emission is depleted within the hot core, it does not address questions pertaining to the source of its emission.

\subsection{If not from hot cores, then where?}\label{subsec:CH2_solution}
\begin{figure}
    \includegraphics[width=1\linewidth]{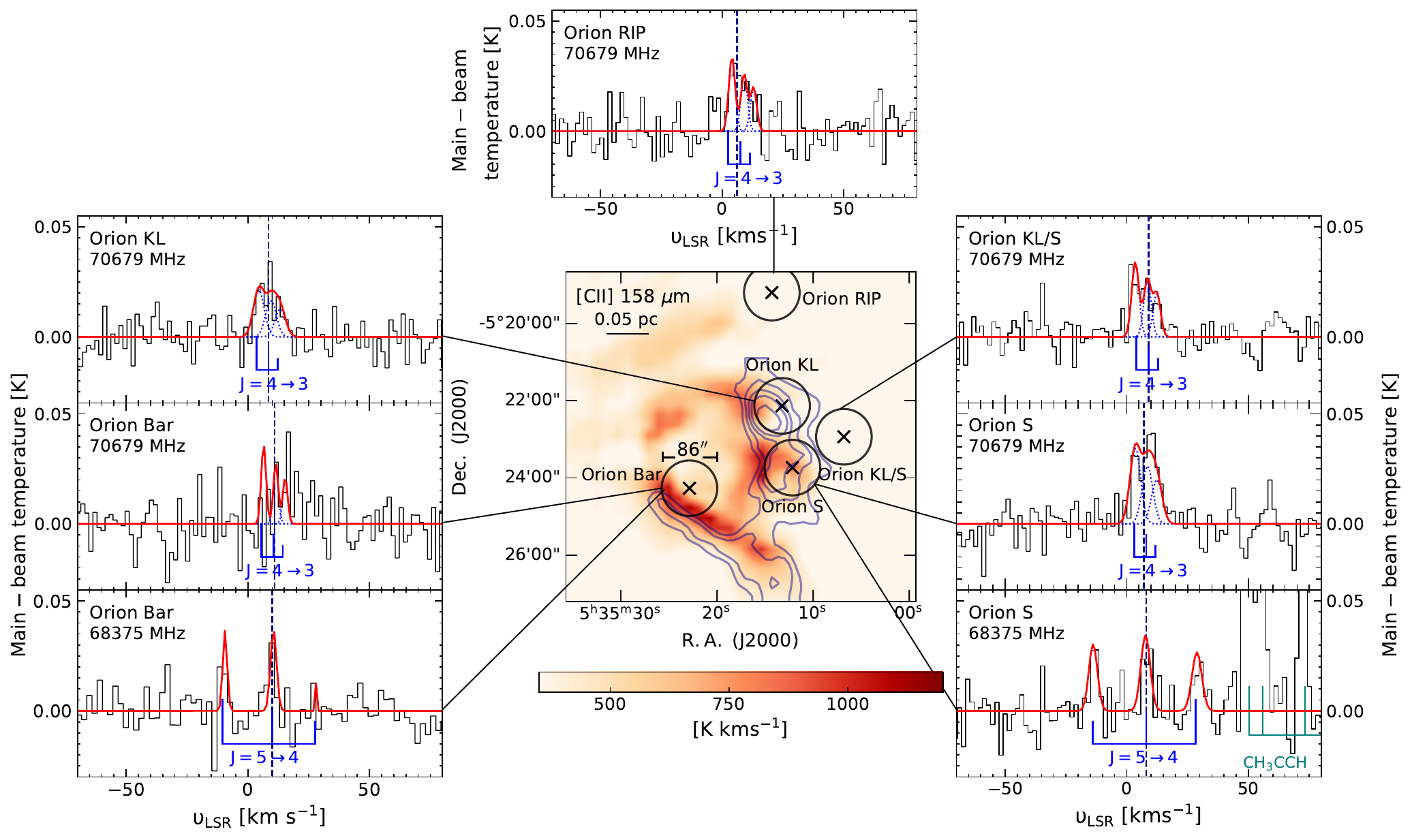}
    \caption{Integrated intensity map of the [C{\tiny II}] 158~$\mu$m emission overlaid with contours of $^{13}$CO in blue from 150 to 400~K~km~s$^{-1}$ in steps of 50~K~km~s$^{-1}$ toward the Orion molecular cloud. The positions observed using the KP 12~m telescope (see text) are marked and labelled by black
crosses with black circles indicating the KP beam. The corresponding CH$_2$ spectra are displayed alongside
the map with the individual HFS and total spectral fits displayed in dotted blue and red curves, respectively. Note that the 68~GHz line observed toward the Orion~S is contaminated by part of the ${K=0~\text{to}~3}$ ladder of the ${J=4\rightarrow3}$ transition of methyl acetylene, CH$_{3}$CCH near 68.3649~GHz. This figure is created based on spectra presented in \citet{Jacob2021CH2} and the C$^+$ and $^{13}$CO maps presented in \citet{Pabst2019} and \citet{Peng2012}, respectively.}
    \label{fig:KP_CH2}
\end{figure}

Motivated by these considerations, \citet{Jacob2021CH2} re-observed the 68-71~GHz lines of CH$_2$ using the KP 12~m telescope toward different positions in the Orion Nebula, including its hot core - Orion KL\footnote{The nominal position toward Orion~KL, corresponds to the peak of the CO molecular line emission and is located at R.A., Dec. = ${05^{\rm h}35^{\rm m}14\rlap{.}{\rm ^{s}}10}, {-05^{\circ}22^{\prime}26\rlap{.}{^{\prime\prime}}54}$.}, Orion South, a position between KL and South, as well as the radical ion peak (RIP) and Orion Nebula's associated photodissociation region (PDR) - the Orion bar. These authors confirm the original detection by \citet{Hollis1995} and report CH$_2$ emission toward all the observed positions with comparable, if not greater line strengths observed toward all the observed positions, with the strongest emission arising from Orion South (see Fig.~\ref{fig:KP_CH2}). In particular, the successful detection of CH$_2$ in emission even toward an intermediate position between Orion~KL and South indicates that its emission is extended and not necessarily confined to the hot core. However, being carried out once again using the KP 12~m telescope, it maybe likely that the signal detected toward the hot-cores likely contain contributions from neighbouring regions located within the 86$^{\prime\prime}$ beam. This necessitated follow-up observations using a moderately sized telescope, to not only verify the extended nature of CH$_2$'s emission but to also reconcile its non-detection when observed using telescopes like the GBT and the IRAM 30~m whose beamsizes are roughly a factor 7 and 3 times smaller than the KP 12~m beam.

\subsubsection{Too big, too small -- just right}
\begin{sidewaysfigure}
\centering
    \includegraphics[width=0.75\textwidth]{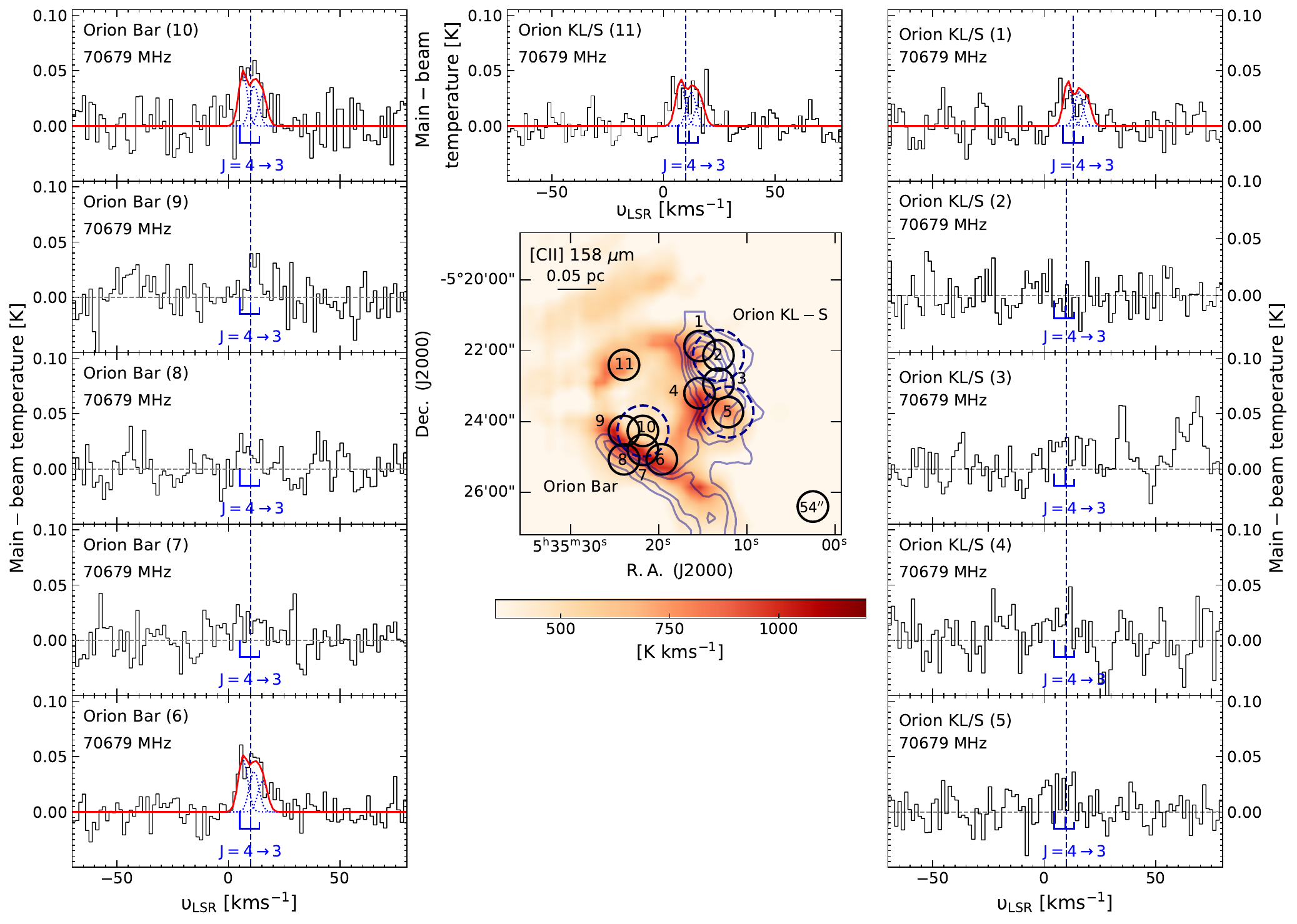}
    \caption{Same as  Fig.~\ref{fig:KP_CH2}, where the pointing positions observed and beam sizes of the Onsala 12~m telescope are numbered and marked in black circles. The dashed blue circles indicated the KP beam toward the nominal Orion~KL, Orion South (S) and Orion Bar positions, respectively, as indicated in Fig.~\ref{fig:KP_CH2}. The corresponding CH$_2$ spectra are displayed alongside
the map with the individual HFS and total spectral fits displayed in dotted blue and red curves, respectively.}
    \label{fig:OSO_CH2}
\end{sidewaysfigure}

To unravel the enigmatic nature of CH$_2$'s emission in the Orion molecular cloud complex, we carried out follow-up observations using the Onsala 20~m telescope which has a FWHM of 54$^{\prime\prime}$ at 70~GHz. Consistent with the findings of the GBT and IRAM observations, \citet{Jacob2021CH2} do not detect any discernible CH$_2$ emission from the nominal position of Orion~KL or Orion South hot-cores above spectral noise levels of on average 22~mK (for a spectral resolution of $\sim 1$~km~s$^{-1}$). To address the question of where the CH$_2$ emission arises from, additional positions in the Orion complex were observed whose positions were chosen based on peaks in the spatial distribution of the [C{\small II}] 158~$\mu$m line emission presented in \citet{Pabst2019}. The successful detection of CH$_2$ emission toward positions where the C$^+$ begins to peak (see Fig.~\ref{fig:OSO_CH2}), the first of which is offset from the nominal Orion KL position by (40\rlap{.}$^{\prime\prime}$2, 24\rlap{.}$^{\prime\prime}$8), now explains the initial detection of CH$_2$ toward the Orion~KL hot core by \citet{Hollis1995} as this position too was covered by the 86$^{\prime\prime}$ beam of the KP 12~m telescope. These results suggest that the CH$_2$ emission is associated with gas layers similar to those traced by [C{\small II}] and resolves the observational discrepancies between the detection and non-detection of CH$_2$ toward the Orion~KL hot core discussed in Sect.~\ref{subsec:CH2_mystery} which also makes certain that the CH$_2$ emission does not originate in compact regions like hot cores. Similarly observations toward the Orion Bar reveal stronger CH$_2$ emission features closer to the more diffuse dissociation front of the PDR. 

\begin{figure}
    \includegraphics[width=1\textwidth]{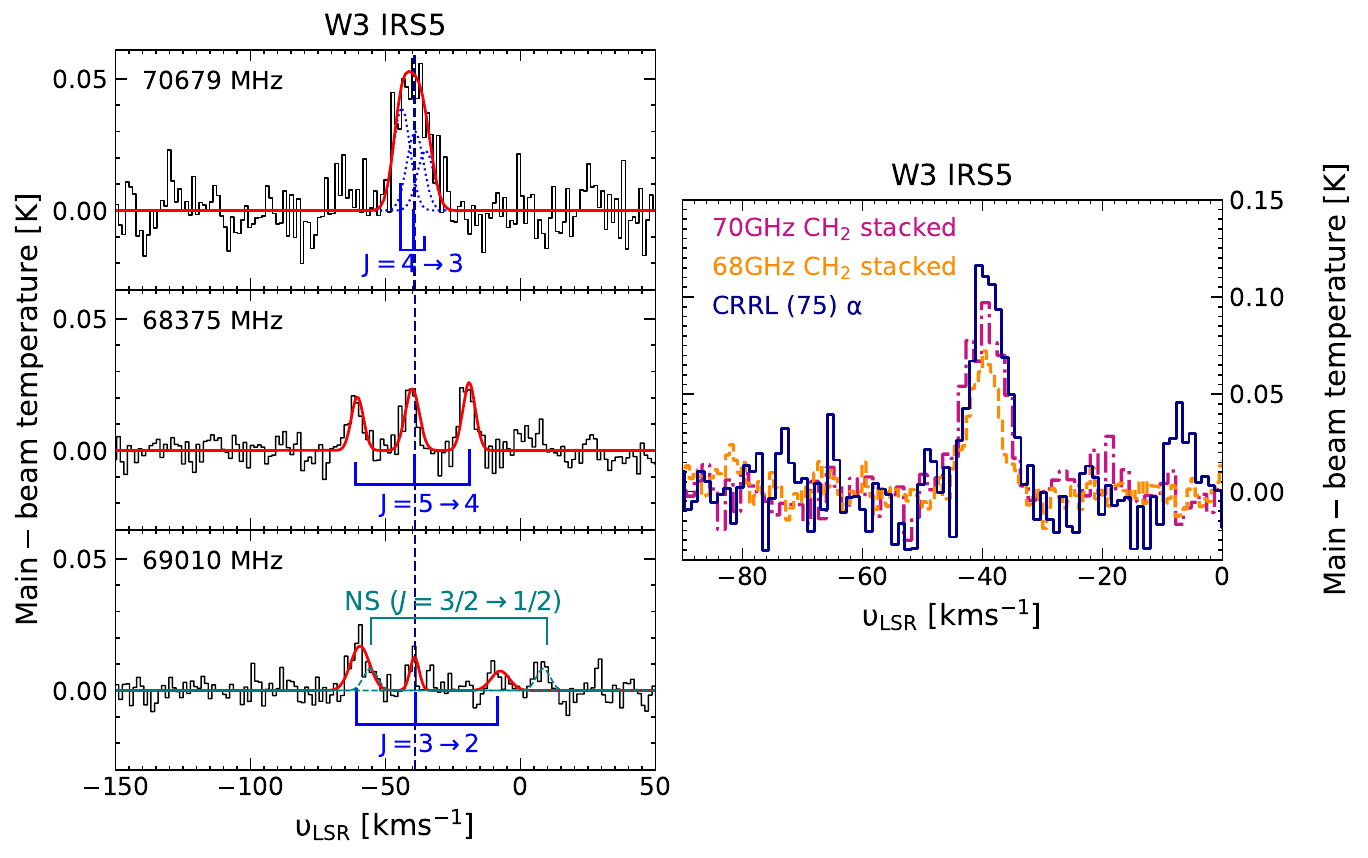}
    \caption{Left: Spectra showing the HFS transitions corresponding to the fine structure levels of CH$_2$ between 68 and 71~GHz observed using the Onsala 20~m telescope toward W3~IRS5. Also marked here are two of the HFS lines of NS (in teal) corresponding to the $J=3/2\rightarrow1/2$ level. The spectra shown in black are overlaid with line fits to individual HFS components and the complete profile using dashed blue and solid red curves, respectively. Right: Decomposed CRRL profile (in blue) alongside the HFS stacked line profiles of the 68~GHz (dashed orange) and 70~GHz (dashed-dotted violet) lines. Adapted based on data presented in \citet{Jacob2021CH2}.}
    \label{fig:W3_example}
\end{figure}

\subsubsection{\texorpdfstring{CH$_2$}{CH2} in other astronomical environments}\label{subsubsec:CH2_other_regions}
Qualitatively, these results point to the origin of the CH$_2$ emission in regions of intermediate gas densities ($n_{\rm H}\sim10^{3}$~cm$^{-3}$ to a few times $10^{4}$~cm$^{-3}$) in the envelopes of hot cores ($T\sim 80$--200~K). In order to confirm the association of the observed CH$_2$ emission with the warm but dilute gas layers of PDRs, the search for CH$_2$ was extended toward well known star-forming regions with associated PDRs. \citet{Jacob2021CH2} successfully detected the (blended) HFS components corresponding to $J=4\rightarrow3$ fine structure line near 70~GHz toward W51~M, N, and E, W49~(N), W3~IRS5, W43, W75~N, DR21, and S140 for the first time. The left-hand panel of Fig.~\ref{fig:W3_example} shows for example the CH$_2$ spectra toward W3~IRS5 toward which all three of the fine-structure components were detected. Remarkably, the line properties (line intensities and widths) of the CH$_2$ lines are comparable with that of carbon radio recombination lines (CRRLs) -- prime tracers of PDRs (see the right-hand panel of Fig.~\ref{fig:W3_example}). This spatial coincidence is particularly useful for determining the physical conditions traced by CH$_2$-bearing gas. Comparing the observed line widths of the two lines and further decomposing them into their thermal ($\Delta\upsilon_{\text{th}}$) and non-thermal ($\Delta\upsilon_{\text{nth}}$) components (Eq.~\ref{eqn:velocity_broadening}), one can estimate the gas temperatures traced by CH$_2$. 
\begin{equation}
        \Delta \upsilon = \sqrt{\Delta\upsilon_{\text{th}}^2 + \Delta\upsilon_{\text{nth}}^2} = \sqrt{\frac{k_{\text{B}}T_{\text{kin}}}{m_\text{C}} + {\left<\Delta\upsilon_{\text{nth}}\right>_{\text{rms}}^2}}
    \label{eqn:velocity_broadening}
\end{equation}
Furthermore, by assuming that both species are impacted by the same turbulent flows and hence will have the same non-thermal broadening, Eq.~\ref{eqn:velocity_broadening} reduces to a simple comparison between the thermal components. For the specific case of W3~IRS5, detailed models studying the physical structure of its PDR gas layers \citep{Sorochenko2000A, Howe1991}, constraints the gas density and temperature near the outer boundaries of the PDR where the C$^+$ and equivalently the CRRL emission peaks, to values of 10$^{5}$~cm$^{-3}$ and 200~K. Therefore, for W3~IRS5 we derive a gas temperature probed by CH$_2$, of at most 233~K. 

While the study by \citet{Jacob2021CH2} exclusively reports the detection of CH$_2$ within PDRs associated with H{\small II} regions they also detail non-detections towards (proto-) planetary nebulae (PNe) and supernova remnants (SNRs). To date, there have also been no reported detections of CH$_2$ toward external galaxies.                                                                                                     

\subsubsection{Physical and excitation conditions}
Non-LTE radiative transfer calculations are essential for a more robust and quantitative analysis of the physical conditions traced. \citet{Jacob2021CH2} carried out such an analysis using the non-LTE radiative transfer code, RADEX \citep{vanderTak2007} facilitated by collisional rate coefficients computed by \citet{Dagdigian2018ch2} for collisions between CH$_2$ and He, later scaled by a factor of 1.4 (or the collision reduced mass) to estimate collisional rate coefficients for collisions of CH$_2$ with H$_2$. Models probed a grid of densities and kinetic temperatures for values of $n_{\rm H_2}$ between 10~cm$^{-3}$ and 10$^6$~cm$^{-3}$ and $T_{\rm kin}$ between 20~K and 300~K, assuming a background temperature of 2.73~K and line width of 5~km~s$^{-1}$, constrained by the observations. The models were further constrained by fixing the column density of CH$_2$ to values of 3$\times 10^{14}$, 5$\times 10^{14}$ and 7$\times 10^{14}$~cm$^{-2}$, obtained by scaling the [CH]/[CH$_2$] ratio determined by \cite{Polehampton2005} with CH column densities derived by \citet{Wiesemeyer2018} toward W~3IRS5. \citet{Jacob2021CH2} find that the observed line intensities and independent line ratios are best reproduced by gas densities and temperatures of 3500~cm$^{-3}$ and 173~K, respectively, albeit with uncertainties in the abundance of CH$_2$. Similar gas densities were derived by \citet{Welty2020}, who report upper limits for the CH$_2$ transition near 1397~\AA\ along the line-of-sight toward HD~62542. The results of the modelling efforts point towards CH$_2$ emission arising from \textit{dilute} PDR layers with densities of a few times 10$^{3}$~cm$^{-3}$ and is consistent with the non-detection of CH$_2$ toward PNe and SNRs, objects which have elevated gas densities of $\sim 10^{5}$--$10^{6}$~cm$^{-3}$ \citep[see for e.g.,][]{Dickman1992, Cox2002}. Similarly, this reconciles the non-detection of CH$_2$ along two lines of sight toward the lens-magnified blazar, PKS 1830$-$211 \citep{Muller2011} where the emitting gas in the foreground galaxy typically hosts densities and temperatures a few times 10$^{3}$~cm$^{-3}$ and 80~K \citep{Schulz2015}, respectively, conditions that may not be hot enough to excite the $N_{K_{\rm a}K_{\rm c}}=4_{04}$--$3_{13}$ lines of CH$_2$.

More interestingly the excitation of these lines reveal weakly negative excitation temperatures $-0.36~$K. This points toward a $\sim1$\% inversion in the population of the $N_{KaKc}$ = $4_{04}-3_{13}$ fine structure transitions of CH$_2$. Since the models reproduced such masing conditions even in the absence of an external radiation field, it suggests that the observed masing in this transition is a robust phenomenon which preferentially populates the upper level of this system. While, the degree of inversion greatly depends on the collisional rate coefficients used and the amount of background pumping available, one can conclude that the weak-masing of this transition plays a role in boosting these lines into detection. However, these conclusions were based on scaling the collisional rate coefficients computed for collisions of CH$_2$ with He to those with H$_2$, an approximation which in a recent review by \citet{vdTak2020} has been shown to be yield low accuracy collisional rate coefficients for collisions with the latter.\\

Recently, fine-, and hyperfine-structure spitting resolved collisional rate coefficients have been computed by \citet{Dagdigian2021} for collisions between \textit{ortho}-, and \textit{para}-CH$_2$ with \textit{ortho-}, and \textit{para-}H$_2$ based on newly computed potential
energy surfaces for the interaction of CH$_2$ in its ground electronic state with H$_2$. \citet{Dagdigian2021} presents a detailed comparison between the newly computed collisional rate coefficients with those for the CH$_2$–H$_2$ collision re-scaled from collisions with He. In general, the collisional rates scaled from He are found to be lower than that computed for collisions with \textit{ortho-}H$_2$ and slightly larger than those computed for collisions with \textit{para-}H$_2$. Furthermore, upon re-running RADEX models with the newly computed collisional rate coefficients, \citet{Dagdigian2021} report line brightness and excitation conditions for the $N_{K_{\rm a}K_{\rm c}}=4_{04}$--$3_{13}$ lines consistent with the values presented in \citet{Dagdigian2018ch2} and \citet{Jacob2021CH2}. 


\subsection{The future of \texorpdfstring{CH$_2$}{CH2} observations}
Although mysteries surrounding the origins of CH$_2$'s emission have been resolved, questions concerning the abundance of this astrochemically important species remain unanswered owing to difficulties in detecting it.\\

The fine-structure lines of the $N_{K_{\rm a}K_{\rm c}}=2_{12}$--$3_{03}$ transitions of \textit{para-}CH$_2$ lying near 444~GHz are amongst the other rotational transitions of this species that are accessible from ground-based observatories. \citet{Jacob2021CH2} were unable to detect this transition down to an rms noise level of 77~mK on average (for a spectral resolution of 1~km~s$^{-1}$) toward the Orion molecular complex with the APEX 12-m sub-millimetre telescope. As noted by these authors, this line remains undetected even with deeper integrations down to a noise level of 8~mK toward selected positions. Non-LTE radiative transfer models carried out by \citet{Jacob2021CH2} and \citet{Dagdigian2021} predict the transition to be seen in absorption with excitation temperatures lower than a few Kelvin up to gas densities $\sim 10^{6}~$cm$^{-3}$ corresponding to very low line brightness temperatures of $\leq 5~$mK. As a result, the chances of detecting significant amounts of \textit{para}-CH$_2$ at 444~GHz are extremely low. \\

Rotational transitions corresponding to the $N_{K_{\rm a}K_{\rm c}} = 1_{11}$--$2_{02}$ \textit{ortho}-CH$_2$ and $N_{K_{\rm a}K_{\rm c}} = 5_{05}$--$4_{14}$ \textit{para-}CH$_2$ levels, lying near 945~GHz and 592~GHz, respectively, also remain undetected in the $480$--$1907$~GHz Herschel/HIFI observations of EXtraOrdinary Sources \citep[HEXOS;][]{Bergin2010} spectral line survey towards Orion South \citep{Tahani2016} and Sgr~B2(M) and (N) \citep{Moller2021}. In addition to the general weakness of the CH$_2$ signal, the non-detection of these lines maybe attributed to a combination of effects including line crowding and confusion at these frequencies. While these high-lying transitions remain undetected, transitions connecting low rotational levels of both spin-states of CH$_2$ lying at FIR wavelengths were observed by \citet{Polehampton2005}, in absorption. Chemically associated with CH (see discussion in Sec.~\ref{subsec:CH2_mystery}), the observed line profiles of the FIR lines of CH$_2$ were fit assuming a constant [CH]/[CH$_2$] ratio, with the best fit attained for a [CH]/[CH$_2$] ratio of 2.7$\pm$0.5. This contradicts the results of early chemical models that predicted CH$_2$ to be the primary product of the dissociative recombination of CH$_3^+$ (90~\%) \citep{van1986comprehensive}, which has since been re-measured in the laboratory revealing branching ratios between CH$_2$, CH, and atomic carbon closer to 35~\%, 30~\% and 35~\% \citep{Vejby1997, Thomas2012}. This difference likely emerges from a more complex chemistry that ensues, which may not have been accounted for in these early models. 
 However, the procurement of additional, higher spectral resolution observations of the FIR lines of CH$_2$ are essential for the accurate assignment of the observed [CH]/[CH$_2$] ratio. A goal which can be achieved using future FIR balloon and/or probe missions. 

\section{The present of hydride studies}\label{sec:the_present}

\subsection{Hydrides at sub-mm and FIR wavelengths}\label{subsec:HyGAL}
\subsection*{SOFIA Legacy program -- HyGAL}
Instruments on the \textit{Herschel} space mission allowed unique access to the ground-state rotational transitions of several interstellar hydrides, a few of which were observed for the first time in the ISM and at high spectral resolution. As discussed in Sect.~\ref{subsec:hydride_history}, the rigorous analysis of their spectra have extended the use of hydrides as diagnostic probes of the different phases of the cold and warm ISM. However, with the end of the \textit{Herschel} mission, it was only SOFIA that provided access to the rotational transitions of hydrides at FIR wavelengths. Having taken timely advantage of this unique facility and its capabilities, the HyGAL (`Hydrides in the GALaxy'; PIs: D. Neufeld, P. Schilke) SOFIA Legacy survey aims to characterise the diffuse Galactic ISM, using absorption line spectroscopic measurements of six key hydrides (ArH$^+$, p-H$_2$O$^+$, OH$^+$, SH, OH, and CH) along with two atomic constituents of the diffuse ISM (C$^+$ and O). Carried out using all modules of the GREAT instrument -- upGREAT and 4GREAT, these observations targeted 25 Galactic sightlines spanning a range of Galactocentric radii between 1~kpc and 10~kpc, selected based on their continuum flux emission at 160~$\mu$m, as estimated from the \textit{Herschel} Hi-GAL compact source catalogue \citep{Elia2021}. For a detailed overview of the project we refer the reader to \citet{Jacob2022}.

Broadly, the HyGAL survey aims to address how molecular clouds -- the building blocks of stars and galaxies-- are formed. As discussed in Sect.~\ref{sec:neutral_ISM}, H\,{\small I} and H$_{2}$ gases form the major components of the neutral ISM and therefore cloud formation must involve the transition from regions dominated by H\,{\small I} gas to those dominated by H$_2$ gas \citep{Federman1979, Sternberg2014, Bialy2016}. While the phase transition from H\,{\small I}-to-H$_{2}$ gas is determined by the environmental conditions in the ambient gas it is primarily driven by macroscopic phenomena that lead to changes in the pressure or density (or column density). Additionally the chemical transition from gas which is mainly atomic to gas which is mainly molecular is a critical step in initiating the growth of chemical complexity in the ISM both locally and across Galactic scales. Theoretical advancements in the field have suggested various macroscopic processes that contribute to the formation of molecular clouds, including gravitational instabilities into spiral arms and the agglomeration of existing clouds \citep[e.g.,][]{Clark2019, Chevance2020MC}. However, there is limited observational evidence to confirm any specific mechanism \citep{Heyer2022}. Statistically sampling a wide range of Galactic sightlines, the data collected under the umbrella of this SOFIA Legacy program combined with comprehensive numerical simulations of diffuse atomic- and molecular clouds will address the following related questions:
\begin{itemize}
    \item What is the distribution of the molecular fraction in different phases of the ISM? A question the HyGAL survey addresses by combining observations of atomic gas tracers like ArH$^+$, and OH$^+$ \citep{Neufeld2010a, Schilke2014, Bialy2019} with molecular gas tracers like CH and OH (discussed in Sect.~\ref{subsec:CH_CO-darkgas}).
    \item  How does the cosmic-ray ionisation rate vary within the Galaxy? Answered by modelling the observed abundances of ArH$^+$, H$_2$O$^+$ and OH$^+$ whose chemistry is driven by the initial ionisation of atoms and molecules by cosmic-rays \citep[see also][]{Indriolo2015, Neufeld2017, Jacob2020ArHp}. 
    \item What is the nature and dissipation of turbulence? As revealed by observations of SH whose formation (unlike the other hydrides discussed here) only takes place via endothermic reactions driven by elevated gas temperatures, ion-neutral drifts in shocks or via turbulent transport \citep{Godard2009, Godard2014, Neufeld2015S, Walch2015}.
\end{itemize} 
In addition to the SOFIA data, these goals are realised with the aid of ancillary data collected using various ground-based facilities like the APEX 12~m sub-mm telescope, the IRAM 30~m telescope, and JVLA. These observations supplement the SOFIA HyGAL data with H{\small I} measurements for abundance estimates (Rugel et al. in prep), as well as observations of sulphur bearing species (like CS and H$_2$S) for the interpretation of the dissipation of turbulence \citep{Neufeld2015S} and other molecular gas tracers like HCO$^+$ and C$_2$H \citep[see,][]{Kim2023}. Furthermore, the JVLA observations simultaneously cover all four of the ground state HFS lines of OH at 18~cm which as discussed in Sect.~\ref{subsec:hydride_history} is often observed to show level inversion. Combined with the FIR rotational transitions of OH observed using SOFIA, the anomalous excitation of the ground state OH lines can be studied similar to that of CH discussed in Sect.~\ref{subsec:CH_maser} by taking advantage of the synergies between its FIR and radio transitions (Busch et al. in prep). 

With the conclusion of the SOFIA mission ${\sim 82}~$\% of the proposed data have been collected under the framework of the HyGAL program. An on-going project, HyGAL plays a significant role in advancing hydride studies specifically in the sub-mm and far-infrared wavelength ranges. This is facilitated by the joint analysis of its different data sets with existing chemical models and developments in numerical simulations of diffuse atomic- and molecular clouds. 

\subsection{Hydrides at radio wavelengths}\label{subsec:radio_wavelengths}

In addition to the OH radio survey described above, observed using the JVLA under the context of the HyGAL Legacy program, there are other large-scale surveys mapping OH in the Milky Way. The largest of which are the Southern Parkes Large-Area Survey in Hydroxyl \citep[SPLASH;][]{Dawson2022}, in the southern hemisphere and The HI/OH/Recombination line survey of the Milky Way \citep[THOR;][]{Rugel2018, Beuther2019} in the northern hemisphere using the JVLA in C-configuration. Both surveys aim to study properties of the cold neutral medium by characterising the kinematic properties and abundances of OH over a large range of hydrogen gas column densities. Furthermore, the complementarity of the two surveys extends beyond their shared goals in that, for those regions in which they both have overlapping observational coverage, the single dish observations under the SPLASH survey (observing extended diffuse emission and absorption against both the compact and extended background continuum) will provide zero spacings for the interferometic data collected under the THOR survey, whereas the THOR survey provides direct optical depth measurements and higher resolution spatial information. However, owing to differences in their sensitivities combining such data sets is non-trivial. The upcoming OH extension of the Galactic Australia Square Kilometre Array Pathfinder survey \citep[GASKAP-OH;][]{Dickey2013}, the MPIfR-MeerKAT Galactic Plane commensal survey \citep[MMGPS;][]{Padmanabh2023} in both L- and S-bands (covering the CH ground state lines), as well as the data collected using the Aericebo telescopes under the Galactic Neutral Opacity and Molecular Excitation Survey \citep[GNOMES;][]{Petzler2020}, all provide additional critical, and complementary data of the Milky Way at radio wavelengths. Toward external galaxies, the same goals are realised within the framework of the on-going Local Group L-Band Survey (LGLBS) targeting six actively star-forming Local Group galaxies in H{\small I} and OH using the JVLA in all four configurations.   

As discussed in Sect.~\ref{subsec:HyGAL}, the analysis of the observed OH lines is not trivial but there is considerable efforts underway to accurately model and understand the properties of its excitation and effects of its masing, both toward the background molecular cloud and along the diffuse sightlines.

\subsection{Hydrides at UV and optical wavelengths}\label{subsec:optical_wavelengths}

In recent years, there has been a shift in the focus of molecular line spectroscopy, with increased attention given to observing the rotational transitions of molecules in the sub-mm and FIR spectral windows. Nevertheless, the search for spectral line signatures across the different wavelength regimes of the electromagnetic spectrum are on-going. In particular, investigations of interstellar molecules via their electronic transitions at UV and optical wavelengths has continued to play an important role in extending our understanding of the diffuse ISM.


While many molecules including hydrides, have electronic transitions that absorb at UV and optical wavelengths, only a handful of molecular species have been detected in the ISM at this wavelength regime, in part due to the effects of dust obscuration. The list of hydrides and hydride isotopologues detected at these wavelengths include, HD \citep{Spitzer1973}, CH \citep{Danks1984, Watson2001}, CH$^+$ \citep{Lambert1986, Crane1995}, OH \citep{Snow1976, Crutcher1976}, OH$^+$ \citep{Krelowski2010, Zhao2015}, SH \citep{Zhao2015SH}, NH \citep{Meyer1991} and HCl \citep{Federman1995}.

Despite the limited spectral resolution achieved by spectroscopic measurements carried out at these wavelengths (relative to that in the sub-mm and FIR regimes), observations of the electronic transitions of hydrides have revealed key molecular associations based on kinematics and abundance measurements. Notably, the CH--H$_{2}$ relationship was first established at UV/optical wavelengths by \citet{Federman1982} and later by \citet{Sheffer2008} and then \citet{Weselak2019} toward stars located in the local diffuse ISM ($10^{19}\,$cm$^{-2} < N(\text{H}_{2})<10^{21}\,$cm$^{-2}$). 
Such an analysis correlating the abundances of CH--H$_2$ (and even CH--CH$^+$) has also been extended to the Magellanic clouds \citep{Welty2006} and supernovae in external galaxies \citep[][and references therein]{Welty2014}. 
Amongst the other hydrides observed, surveys of the electronic transitions of OH and OH$^+$ have established significant correlations of these species with molecular \citep{Porras2014, Mookerjea2016, Weselak2020, Rawlins2023}, and atomic gas \citep{Porras2014, Weselak2020}, respectively, further emphasising their diagnostic capabilities. Recent observations by \citet{Bacalla2019}, of the electronic bands of OH$^+$ carried out using VLT-UVES under the framework of the ESO Diffuse Interstellar Bands Large Exploration Survey \citep[EDIBLES;][]{Cami2018} were used to determine the cosmic-ray ionisation rates. These authors reported cosmic-ray ionisation rates between (3.9--16.4)$\times 10^{-16}$~s$^{-1}$, values that are higher than those typically derived using previous measurements of OH$^+$ at sub-mm and FIR wavelengths \citep{Indriolo2015, Neufeld2017, Jacob2020ArHp}.\\

From the discussions thus far, it is clear that many of the hydride species discussed above have been detected via both their electronic transitions at UV/optical wavelengths as well as through their rotational transitions at sub-mm/FIR wavelengths. The careful combination of these multi-wavelength observations alongside that of other interstellar species carried out toward same clouds can potentially provide detailed information on the chemistry of these clouds, extend kinematic correspondences and help develop a more comprehensive picture of diffuse gas environments \citep{Crutcher1985, Gredel1994, Rice2018}. This emphasises the need to revisit and attain more observations of the electronic transitions of hydrides at UV and optical bands. Moreover, there is yet more to be explored within this spectral window, including the identification of spectral line signatures of hydrides such as H$_2$O and HF, amongst others \citep{Snow2005}, that have eluded detection despite their large predicted abundances in early diffuse cloud models, at these wavelengths \citep{van1986comprehensive}. 

\section{Summary and the future of hydride studies}\label{sec:summary_outlook}
A tale of hydrides past, present, and future, this review has focused on the particular use of carbon hydrides CH and CH$_2$ alongside CH isotopologues ($^{13}$CH, CD) and other related species like OH, as diagnostics for the different phases of the ISM.
This work summaries the following results:
\begin{enumerate}
    \item Large fractions of molecular material have been identified as residing in CO-dark gas and is widely studied using C$^+$ as a proxy for H$_2$. However, it is necessary to calibrate the contributions of C$^+$ because it can also arise in other phases of the ISM. Demonstrating the effectiveness of CH as a tracer for CO-dark gas across Galactic scales highlights its use as a complementary tracer to C$^+$. The utility of which should be extended toward studies of external galaxies. 
    
    \item Through a comprehensive understanding of the physical and excitation conditions traced by the extensively observed ground state lines of CH at a wavelength of 9~cm, previous limitations on its usage as a tracer are now being resolved.   
    \item The detection of $^{13}$CH presents an unbiased tracer for the $^{12}$C/$^{13}$C ratio-- an important diagnostic for probing the nucleosynthesis history of a region while that of CD provides direct evidence for deuterated pathways proceeding via chemical pathways involving CH$_2$D$^+$ or warm deuteration. As progenitor species responsible for the 13-C and D atom substitutions in carbon-bearing molecules subsequently formed, they fill crucial gaps in our understanding of the underlying chemistry and effects such as fractionation. 
    
    \item 
    By successfully unravelling the enigmatic emission of interstellar CH$_2$, a satisfactory conclusion wherein both theory and observations agree, has been reached. This conclusion indicates that CH$_2$'s emission originates from warm yet dilute gas layers in PDRs, with its abundance peaking at the edges of dense clouds similar to CH. 
\end{enumerate}
HyGAL promises a valuable legacy for future studies of the ISM, extending the big impact of these small molecules but a big question that looms is \textit{what is the future of hydride studies?}\\

While the immediate future lacks high resolution air-, and space borne instrumentation probing FIR wavelengths, design plans are underway for prospective long duration balloon missions and FIR probe class missions. But as discussed above the observations of hydrides are not limited to this wavelength regime and the future of hydride studies will continue to take advantage of the increasing capabilities of new and existing facilities operating across the electromagnetic spectrum from UV/optical to FIR wavelengths. 
  
At sub-mm wavelengths, this includes observations of the high-lying rotational transitions of hydrides that are accessible from the ground using facilities like the APEX, the 12-m Caltech Sub-millimetre telescope (CSO), ALMA and the NOrthern Extended Millimetre Array (NOEMA). The latter also facilitates high-resolution hydride observations toward high-redshift galaxies. Despite the richness of the sub-millimetre window, some species are best observed in the near-, and mid-infrared wavelength regime accessible with NASA's Infrared Telescope Facility (IRTF) and now JWST. This includes the observations of the rovibrational transitions of hydrides including HeH$^+$ and CH$^+$ \citep{Neufeld2020, Neufeld2021}, as well as observations of methane and water ices amongst that of other radicals \citep[see for e.g.,][]{McClure2023}. In addition, the higher resolution currently achieved in this wavelength window also presents with it a new detection space that has already reported the recent tentative detection of CH$_3^+$ (Berne et al. 2023).

In the radio regime constraints on the excitation conditions of the ground state HFS lines of CH presented in \citet{Jacob2021} and OH described within the context of the HyGAL mission and in Sect.~\ref{subsec:radio_wavelengths}, will firmly establish these widely, and relatively easily observable lines as powerful probes of the diffuse skies at radio-wavelengths. An analysis that can be invoked even without accurate column density measurements using the masing action of well-studied sources as templates for those that lack FIR data. In addition, both single dish as well as interferometric facilities like the JVLA for northern sources and the MeerKAT array and ASKAP for the southern sources are well suited to observe these lines, which when mapped can help extend our understanding of the nature of CO-dark gas (see Sect.~\ref{subsec:CH_CO-darkgas}, ~\ref{subsec:HyGAL}). The next generation of, and upgrades to radio astronomy facilities will see the addition of phased array feeds (PAFs) which will extend the observational parameter space with larger collecting areas and fields of view. With respect to hydride observations, the greater field of view provided by PAFs will open up new possibilities for efficiently mapping large sky surveys of the (relatively) weaker and more extended emission and absorption features of CH and OH. Moreover, wider collecting areas will also enable us to probe deeper into the Universe.

In the preceding sections a variety of modelling techniques utilising experimental and quantum chemical calculations have been employed to analyse the hydride observations discussed. Therefore, as noted earlier, progressing our understanding of the signatures observed, developments in the field of laboratory astrophysics, and advancements in reproducing robust chemical and physical models. To exemplify the significance of these collaborative studies, we briefly discuss recent findings related to CH$^+$. In the laboratory, \citet{Paul2022} have measured the reaction rates for the dissociative recombination (DR) of CH$^+$ with electrons achieved using Cryogenic Storage Ring (CSR) experiments in Heidelberg. The measurements reported by these authors significantly reduce the uncertainties in the DR rate for CH$^+$, from over an order of magnitude to $\sim$20\%. In the context of astrophysics, this provides new constraints on the physical regime in which DR reactions form the dominant destruction pathway for CH$^+$. Similar experiments are currently being developed at the CSR to measure the DR of other key ions including that of OH$^+$. By running 3D simulations of the diffuse and turbulent multi-phase ISM, \citet{Godard2023} have built a coherent picture for the formation of CH$^+$-- an otherwise longstanding puzzle. Their findings point towards the formation of CH$^+$ resulting from the exchange of matter between the cold and warm phases of the neutral ISM driven by turbulent advection and thermal instabilities. In addition to presenting a new solution for the origins of CH$^+$, the two-phase turbulent mixing producing CH$^+$ hints to it tracing the H$_2$ mass loss rate of cold neutral medium clouds. In the future we can look forward to extensions of such an analysis carried out toward other chemical species including other hydrides.  

Overall the future of hydride studies appears promising, with exciting new prospects for multi-wavelength observations, advanced laboratory measurements, and quantum chemical calculations, and increasingly detailed and dedicated models to look forward to. 

\backmatter

\bmhead{Acknowledgements}
{A.M.J is grateful to the anonymous referee, for a careful review of the manuscript of this article and their insightful comments which have helped to improved this manuscript. A.M.J would like to thank Karl M. Menten and David Neufeld for their support and guidance and Paule Sonnentrucker, Michael Busch and Michael R. Rugel for their comments on specific sections of this summary. Part of this research was carried out when A.M.J. was a member of the International Max Planck Research School (IMPRS) for Astronomy and Astrophysics at the Universities of Bonn and Cologne. A.M.J. was generously supported by USRA through a grant for SOFIA Program 08-0038.}

\section*{Declarations}
\begin{itemize}
    \item Funding \\
    A.M.J. was generously supported by USRA through a grant for SOFIA Program 08-0038. \\
    \item Conflict of interest\\
    The authors have no competing interests to declare
that are relevant to the content of this article.
    \item Availability of data and materials\\
    All data generated or analysed during this study are included in this published article (and its supplementary information files).
\end{itemize}

\newpage
\begin{appendices}

\section{Spectroscopic parameters}\label{appendix:frequencies}
This Appendix presents the frequencies and relevant spectroscopic parameters of the different transitions discussed in this work for CH and its isotopologues in Tables~\ref{tab:CH-spec_properties} and \ref{tab:CH-isotopes-spec_properties}, and CH$_2$ in Table~\ref{tab:CH2-spec_properties}, respectively. 

\begin{table}[!ht]
    \begin{center}
    \caption{Spectroscopic properties of the hyperfine-structure splitting transitions of CH. The columns are (from left to right): the transition as described by the hyperfine quantum number ($F$), the frequency of the transition and its reference, the Einstein A coefficient ($A_{\rm ul}$) and the upper level energy of the transition ($E_{\rm u}$), alongside the telescope/array and receiver used to carry out these observations.}
    \begin{tabular}{rrlccc}
    \hline \hline 
       Transition  & \multicolumn{1}{c}{Frequency\footnotemark[2]} & Ref &  $A_{\rm ul}$\footnotemark[1] &  $E_{\rm u}$ & Telescope/ \\
        $F^{\prime}$--$F^{\prime\prime}$ & \multicolumn{1}{c}{[MHz]} &  & ~[s$^{-1}$] & [K] & Receiver \\ 
         \hline 
         \multicolumn{6}{c}{$N, J = 1, 1/2$} \\
         \hline
        $0^{-}$--$1^{+}$ & 3263.793(3) & [1] & 2.876($-$10) & 0.16 & JVLA \\
        $1^{-}$--$1^{+}$ & 3335.479(3) & & 2.045($-$10) & \\
        $1^{-}$--$0^{+}$ & 3349.192(3) & & 1.036($-$10) &  \\
         \hline 
          \multicolumn{6}{c}{$N, J = 1, 3/2 \rightarrow 1, 1/2$} \\
         \hline
     $1^{-}$--$1^{+}$ & 532721.588(2) &  [2] & 2.069($-$4) & 25.72  & \textit{Herschel}/HIFI\\
     $2^{-}$--$1^{+}$ & 532723.889(1) &  & 6.207($-$4) \\
     $1^{-}$--$0^{+}$ & 532793.274(1) &  & 4.139($-$4) \\
     $2^{+}$--$1^{-}$ & 536761.046(1) &  & 6.378($-$4) & 25.76 \\
     $1^{+}$--$1^{-}$ & 536781.856(1) &  & 2.126($-$4)  \\
     $1^{+}$--$0^{-}$ & 536795.569(1) &  & 4.251($-$4)  \\
           \hline 
          \multicolumn{6}{c}{$N, J = 2, 3/2 \rightarrow 1, 1/2$} \\
         \hline
  $1^{-}$--$1^{+}$ & 2006748.918(64) & [3]& 1.117($-$2) & 96.31 & SOFIA/upGREAT\\
                    $1^{-}$--$0^{+}$ & 2006762.630(64) & & 2.234($-$2)   \\
					$2^{-}$--$1^{+}$ & 2006799.118(64) & & 3.350($-$2) \\
 \footnotemark[7]$1^{+}$--$1^{-}$& 2010738.641(64) & & 1.128($-$2) & 96.66\\
                    \footnotemark[7]$1^{+}$--$0^{-}$ & 2010810.460(100) & & 2.257($-$2) &  \\				 
                    \footnotemark[7]$2^{+}$--$1^{-}$ & 2010811.920(100) & & 3.385($-$2) &  \\
\hline  
\label{tab:CH-spec_properties}
    \end{tabular}
    \end{center}
 \footnotetext{The frequencies and other spectroscopic parameters are taken from [1]:~\citet{Truppe2014}, [2]:~\citet{Amano2000}, [3]:~\citet{Davidson2001} and the Cologne Database for Molecular Spectroscopy \citep[CDMS;][]{Endres2016}. $^{(\dagger)}$: Values in parenthesis represent uncertainties in the rest frequencies of the hyperfine-structure splitting lines, in units of the last significant digits. $^{(*)}$: The exponential term is presented inside the parenthesis. $^{(**)}$: These transitions are not observed since they are contaminated by ozone absorption features at 149.1558~$\mu$m and 149.7208~$\mu$m that originate from the signal band.}
    
\end{table}
\begin{table}[h!]
    \begin{center}
    \caption{Same as Table~\ref{tab:CH-spec_properties} but presents the spectroscopic properties for hyperfine-splitting structure transitions of $^{13}$CH and CD.}
    \begin{tabular}{rrlccc}
    \hline \hline 
       Transition  & \multicolumn{1}{c}{Frequency\footnotemark[2]} & Ref &  $A_{\rm ul}$\footnotemark[1] &  $E_{\rm u}$ & Telescope/ \\
        $F^{\prime}$--$F^{\prime\prime}$ & \multicolumn{1}{c}{[MHz]} &  & [s$^{-1}$] & [K] & Receiver \\ 
         \hline 
         \multicolumn{6}{c}{$^{13}$CH: $N, J = 2,3/2 \rightarrow 1, 1/2$} \\
         \hline
    $3/2^{+}$--$1/2^{-}$ & 1997423.2(40) & [1] & 1.145(-3) & & SOFIA/upGREAT \\
    $3/2^{+}$--$3/2^{-}$ & 1997446.4(40) & & 0.573(-3) & & \\
    $5/2^{+}$--$3/2^{-}$ & 1997443.7(40) & & 1.909(-3) & & \\
    $3/2^{-}$--$1/2^{+}$ & 2001567.2(40) & & 1.157(-3) & &\\
    $3/2^{-}$--$3/2^{+}$ & 2001223.0(40) & & 0.578(-3)\\
    $5/2^{-}$--$3/2^{+}$ & 2001367.3(40) & & 1.929(-3)\\ 
         \hline 
          \multicolumn{6}{c}{CD: $N, J = 1, 3/2 \rightarrow 1, 1/2$} \\
         \hline
        $1/2^{+}$--$3/2^{-}$ & 439254.774(45) & [2] & 0.555(-4) & 23.07 & APEX/nFLASH460 \\
        $3/2^{+}$--$3/2^{-}$ & 439255.608(30) &  & 2.223(-4) & & \\
        $5/2^{+}$--$3/2^{-}$ & 439257.450(30) &  & 5.003(-4) & & \\
        $1/2^{+}$--$1/2^{-}$ & 439271.905(30) &  & 4.447(-4) & &\\    $3/2^{+}$--$1/2^{-}$ & 439272.694(30) &  & 2.779(-4) & & \\
        $5/2^{-}$--$3/2^{+}$ & 439794.923(30) &  & 5.033(-4) & 23.04 & \\
        $3/2^{-}$--$3/2^{+}$ & 439800.005(30) &  & 2.236(-4) & & \\
        $3/2^{-}$--$1/2^{+}$ & 439803.008(30) &  & 2.796(-4) & & \\
        $1/2^{-}$--$3/2^{+}$ & 439803.124(45) &  & 0.559(-4) & & \\
        $1/2^{-}$--$1/2^{+}$ & 439806.093(30) &  & 4.473(-4) & & \\
           \hline 
          \multicolumn{6}{c}{CD: $N, J = 2, 3/2 \rightarrow 1, 1/2$} \\
         \hline
     \footnotemark[7]$1/2^{-}$--$3/2^{+}$ & 1325248.368(276) &   &  0.608(-3)  & 65.54 & SOFIA/4GREAT\\
     \footnotemark[7]$1/2^{-}$--$1/2^{+}$ & 1325251.364(278) &  & 4.869(-3) & & \\
     \footnotemark[7]$3/2^{-}$--$3/2^{+}$ & 1325255.172(210) &  & 2.434(-3) & & \\
     \footnotemark[7]$3/2^{-}$--$1/2^{+}$ & 1325258.167(213) &  & 3.043(-3) & & \\
     \footnotemark[7]$5/2^{-}$--$3/2^{+}$ & 1325266.217(194) &  & 5.477(-3) & & \\
     \footnotemark[7]$1/2^{+}$--$3/2^{-}$ & 1325775.667(261) &  & 0.613(-3) & 65.62 & \\
     \footnotemark[7]$3/2^{+}$--$3/2^{-}$ & 1325784.078(197) &  & 2.451(-3) & & \\
     \footnotemark[7]$1/2^{+}$--$1/2^{-}$ & 1325792.765(258) &  & 4.902(-3) & & \\
     \footnotemark[7]$5/2^{+}$--$3/2^{-}$ & 1325797.812(194) &  & 5.515(-3) & & \\
     \footnotemark[7]$3/2^{+}$--$1/2^{-}$ & 1325801.181(195) &  & 3.064(-3) & & \\
\hline  
 \label{tab:CH-isotopes-spec_properties}
    \end{tabular}
    \end{center}
 \footnotetext{The frequencies and other spectroscopic parameters are taken from [1]:~\citet{Davidson2004}, [2]:~\citet{Halfen2008}, and 
 the Jet Propulsion Laboratory \citep[][JPL;]{Pickett1998} database. $(\dagger)$: Values in parenthesis represent uncertainties in the rest frequencies of the hyperfine-structure splitting lines, in units of the last significant digits. $^{(*)}$: The exponential term is presented inside the parenthesis. $^{(**)}$: These lines remain undetected down to a noise level of 0.55~K (at a spectral resolution of 0.1~km~s$^{-1}$) toward star-forming region W49~N \citep[see][for more details]{Jacob2023}.}
   
\end{table}
\newpage
\begin{table}[h!]
    \begin{center}
    \caption{Same as Table~\ref{tab:CH-spec_properties} but presents the spectroscopic properties for hyperfine-splitting structure transitions of CH$_{2}$.}
    \begin{tabular}{llllccc}
    \hline \hline 
       \multicolumn{2}{c}{Transition}  & \multicolumn{1}{c}{Frequency\footnotemark[2]} & Ref &  $A_{\rm ul}$\footnotemark[1] &  $E_{\rm u}$ & Telescope/ \\
       $J^{\prime}$--$J^{\prime\prime}$ & $F^{\prime}$--$F^{\prime\prime}$ & \multicolumn{1}{c}{[MHz]} &  & [s$^{-1}$] & [K] & Receiver \\ 
         \hline 
         \multicolumn{7}{c}{$N_{K_{\text{a}}K_{\text{c}}}  = 2_{12}-3_{03}$ (para-CH$_{2}$)} \\
         \hline
         3--4 & -- & 444825.666(30) & [1] & 5.977(-5)
          & 155.97 & APEX/FLASH460\\
         2--3 & -- & 439960.991(30) & & 6.432(-5)
          & 156.27\\
         1--2 & -- & 444913.930(30) & & 7.006(-5)
         & 155.85\\
         \hline 
         \multicolumn{7}{c}{$N_{K_{\text{a}}K_{\text{c}}} = 4_{04}-3_{13}$ (ortho-CH$_{2}$)} \\
         \hline 
        5--4 & 6--5 & 68371.278(41) & [2] & 2.165(-7) & 224.22 & Onsala/4-mm Rx.\\
        & 5--4 & 68375.875(39) &  & 2.079(-7) & &\\
        & 4--3 & 68380.873(41) &  & 2.060(-7) & &\\ 
        4--3 & 3--2 & 70678.633(42) & & 2.053(-7) & 224.76 & \\
        & 4--3 & 70679.543(45) & & 2.097(-7) & &\\
        & 5--4 & 70680.720(38) & & 2.236(-7) & &\\ 
        3--2 & 2--1 & 69007.179(74) & & 1.717(-7) &  224.15 & \\
        & 3--2 & 69014.202(37) & & 1.817(-7) & &\\
        & 4--3 & 69019.187(44) & & 2.044(-7) & &\\ 
\hline  
\label{tab:CH2-spec_properties}
    \end{tabular}
    \end{center}
 \footnotetext{The frequencies and other spectroscopic parameters are taken from [1]:~\citet{Brunken2005}, [2]:~\citet{Lovas1983}, \citet{Ozeki1995} and the Cologne Database for Molecular Spectroscopy \citep[CDMS;][]{Endres2016}. $\dagger$: Values in parenthesis represent uncertainties in the rest frequencies of the hyperfine-structure splitting lines, in units of the last significant digits. $^{(*)}$: The exponential term is presented inside the parenthesis. $^{**}$: These transitions remain undetected down to a noise level of 8~mK (at a spectral resolution of 1~km~s$^{-1}$) toward multiple positions in the Orion molecular cloud.}
    
\end{table}




\end{appendices}

\newpage

\bibliography{sn-bibliography}


\end{document}